\PassOptionsToPackage{table,xcdraw}{xcolor}
\documentclass[acmsmall]{acmart}

\AtBeginDocument{%
  }

\setcopyright{rightsretained}
\acmDOI{10.1145/3660786}
\acmYear{2024}
\copyrightyear{2024}
\acmSubmissionID{fse24main-p241-p}
\acmJournal{PACMSE}
\acmVolume{1}
\acmNumber{FSE}
\acmArticle{79}
\acmMonth{7}
\received{2023-09-28}
\received[accepted]{2024-04-16}

\usepackage[normalem]{ulem}
\usepackage{macros}
\usepackage{multirow}
\usepackage{adjustbox}
\PassOptionsToPackage{hyphens}{url}
\usepackage{hyperref}
\usepackage{xurl}
\usepackage{tcolorbox}
\usepackage[]{mdframed}

\usepackage{algorithm}
\usepackage[noend]{algpseudocode}

\usepackage{enumitem}

\usepackage{array}
\newcolumntype{R}[2]{%
    >{\adjustbox{angle=#1,lap=\width-(#2)}\bgroup}%
    l%
    <{\egroup}%
}
\newcommand{\rot}[1]{\adjustbox{angle=80,lap=\width-1em}{#1}}
\newcommand{\vertical}[1]{\adjustbox{angle=90,lap=\width-1em}{#1}}

\usepackage{xspace}

\usepackage{cleveref}
\usepackage{hhline}

\usepackage{graphicx}
\usepackage{subcaption}

\usepackage{pifont}
\newcommand{\xmark}{\ding{55}}%

\usepackage[table]{xcolor}
\usepackage{soul}

\usepackage{booktabs} 

\newcommand{\revise}[1]{\textcolor{black}{#1}}
\newcommand{\camera}[1]{\textcolor{black}{#1}}
\begin{document}



\title{Demystifying Invariant Effectiveness for Securing Smart Contracts (Extended Version)}


\author{Zhiyang Chen}
\orcid{0000-0002-2315-397X}
\affiliation{%
  \institution{University of Toronto}
  \city{Toronto}
  \country{Canada}
}
\email{zhiychen@cs.toronto.edu}

\author{Ye Liu}
\orcid{0000-0001-6709-3721}
\affiliation{%
  \institution{Nanyang Technological University}
  \city{Singapore}
  \country{Singapore}
}
\email{ye.liu@ntu.edu.sg}

\author{Sidi Mohamed Beillahi}
\orcid{0000-0001-6526-9295}
\affiliation{%
  \institution{University of Toronto}
  \city{Toronto}
  \country{Canada}
}
\email{sm.beillahi@utoronto.ca}

\author{Yi Li}
\orcid{0000-0003-4562-8208}
\affiliation{%
  \institution{Nanyang Technological University}
  \city{Singapore}
  \country{Singapore}
}
\email{yi_li@ntu.edu.sg}

\author{Fan Long}
\orcid{0000-0001-7973-1188}
\affiliation{%
  \institution{University of Toronto}
  \city{Toronto}
  \country{Canada}
}
\email{fanl@cs.toronto.edu}


\newcommand*{\name}{{\textsc{Trace2Inv}}\xspace}

\begin{abstract}
  Smart contract transactions associated with security attacks often exhibit distinct behavioral
  patterns compared with historical benign transactions before the attacking events.
  While many runtime monitoring and guarding mechanisms have been proposed to validate invariants
  and stop anomalous transactions on the fly, the empirical effectiveness of the invariants used
  remains largely unexplored.
  In this paper, we studied $23$ prevalent invariants of $8$ categories, which are either deployed in high-profile
  protocols or endorsed by leading auditing firms and security experts.
  Using these well-established invariants as templates, we developed a tool {\name} which dynamically
  generates new invariants customized for a given contract based on its historical transaction data.
  We evaluated {\name} on $42$ smart contracts that fell victim to $27$ distinct exploits on the
  Ethereum blockchain.
  Our findings reveal that the most effective invariant guard alone can
    successfully block $18$ of the $27$ identified exploits with minimal gas
    overhead. Our analysis also shows that most of the invariants remain
    effective even when the experienced attackers attempt to bypass them.
  Additionally, we studied the possibility of combining multiple invariant
    guards, resulting in blocking up to $23$ of the $27$ benchmark exploits and
    achieving false positive rates as low as $0.28\%$. \camera{ {\name} 
    significantly outperforms state-of-the-art works on smart contract invariant mining and transaction attack 
    detection in accuracy. {\name} also surprisingly found two previously unreported
    exploit transactions.} 

\end{abstract}




\maketitle

\vspace{-2ex}
\section{Introduction}
\vspace{-1ex}

Blockchain technology has paved the way for decentralized, resilient, and
programmable ledgers on a global scale. One of its most impactful applications
is smart contracts. These smart
contracts allow developers to encode intricate transaction rules that govern
the ledger. This innovation has made both blockchains and smart contracts
essential infrastructure for decentralized financial services,
commonly known as DeFi. As of Sept 25th, 2023, the Total digital asset Value Locked (TVL) in
$2,933$ DeFi protocols has reached an impressive $48.58$
billion~\cite{defillama2}.

However, the landscape is not without its challenges. Security attacks pose a
significant threat to the security of smart contracts. Attackers can exploit
various vulnerabilities by sending malicious transactions, potentially leading
to the theft of millions of dollars. As of Sept 25th, 2023, the financial
losses attributed to security attacks on DeFi protocols exceeded $5.53$ billion
USD~\cite{defillama}.

One key observation is that transactions initiated by attackers often display
abnormal behaviors when compared to standard transactions from regular DeFi
contract users. These malicious transactions may exploit control flows in
corner cases, use abnormally large values to trigger overflows, or manipulate a
large volume of digital assets to distort the market in DeFi contracts. In
fact, industry experts have been actively monitoring abnormal digital asset
movements on-chain to report malicious activities. For example, Forta Network~\cite{Forta}
deploys monitoring bots to detect on-chain security-related events in real-time. 
Driven by this observation, smart contract developers have proposed deploying
runtime checks to detect transactions leading to abnormal behaviors to neutralize
malicious attacks. These checks involve enforcing various runtime invariants,
such as restricting the maximum number of digital asset deposits or withdrawals
in a contract to prevent market manipulation. Another example is to limit the
interaction between contracts to prevent attackers from crafting sophisticated
attack strategies. However, these mechanisms are often manually designed and
tailored for specific DeFi protocols. This raises questions about their
effectiveness across different types of contracts and whether they maintain an
acceptable false positive rate without hindering normal user activities.

\noindent \textbf{Smart Contract Invariant Study:}
This paper presents the first comprehensive, quantitative analysis focused on
the utilization of dynamically inferred invariants to enhance smart contract security. We
examine $23$ invariant templates, which are advocated by leading auditing firms,
academic research, and DeFi protocol developers. Our findings indicate that
dynamically inferred dynamic invariants serve as effective mechanisms for 
thwarting security
breaches. This paper then proposes new strategies to combine multiple invariants effectively.
Our combined invariants can neutralize over
74.1\% of malicious attacks while maintaining a false positive rate of less than
0.28\%.

\noindent \textbf{{\name}:}
To facilitate this study, we have developed {\name}, a scalable and extensible
invariant synthesis framework. {\name} is designed to automatically derive
invariants from transaction traces through the use of trace and dynamic taint analysis. 
{\name} leverages the main feature of public blockchains, 
transparent databases of transactions histories containing well-organized transaction execution data. 
Then, for each invariant template under consideration, {\name} employs a specialized inference
algorithm to dynamically generate the corresponding invariant based on
historical transaction data. 

\noindent \textbf{Experimental Results:}
We evaluate {\name} on a benchmark set of $42$ smart contracts that have
previously fallen victim to security attacks. Our results show that properly
constructed invariants are effective in neutralizing security threats in $39$
out of the $42$ benchmark contracts. 

In the course of our study, we categorized the $23$ invariant templates into
eight distinct groups based on their underlying design principles: access
control, time lock, gas control, re-entrancy, oracle, storage, money flow, 
and data flow. Subsequently, we conducted a series of in-depth analyses to compare
the efficacy of invariants within each group. We also manually scrutinized the
transactions flagged by each invariant template, leading to several key
findings:
\vspace{-1mm}
\begin{itemize}[leftmargin=*]
\item \textbf{Finding 1:} Certain invariant outperform others in
    terms of effectiveness. Within each invariant group, we identified at least one
        invariant template that is quantitatively superior, neutralizing a
        greater number of attacks while generating fewer false positives.
        See Section~\ref{sec:rq1}.

\item \textbf{Finding 2:} Invariants remain effective even when attackers are
    aware of them in the majority of cases. A common concern regarding runtime
        invariants is their potential vulnerability to informed attackers. Our
        study reveals that selected invariants in the access control, time lock,
        gas control,
        money flow, and data flow groups often directly counter critical elements of
        attack strategies, such as flash loans and transaction atomicity. These
        invariants not only neutralize the malicious transactions but also
        render the attack strategies unfeasible or non-profitable in 84.21\% of
        cases. See Section~\ref{sec:rq2}.

\item \textbf{Finding 3:} Normal users can possibly circumvent invariant guards, 
        thereby mitigating the impact on user
        experience. For example, in the case of data flow and money flow
        invariants, a user can divide a large transaction into smaller segments
        to bypass the invariant guard in 80\% false positive instances. 
        See Section~\ref{sec:rq2}.

\item \textbf{Finding 4:} Combined invariants, formed through disjunction or
    conjunction, offer enhanced security coverage with lower false positive rate.
    Different groups of
        invariants address different attack scenarios. A combined invariant
        formed through conjunction can cover more attack vectors, while one
        formed through disjunction may reduce the false positive rate, as
        malicious attacks often exhibit multiple abnormal behaviors. See
        Section~\ref{sec:rq3}.
\end{itemize}
\noindent \textbf{Contributions:} This paper presents the following contributions:
\begin{itemize}[leftmargin=*]
    \item \textbf{Invariant Inference:} This paper conducts an extensive study
        of $23$ invariant templates, categorized into $8$ distinct groups.
        Additionally, we introduce innovative techniques for the effective
        inference of invariants across all studied templates from transaction 
        history. 

    \item \textbf{{\name}:} This paper presents the design and implementation of {\name},
        a specialized tool for smart contract trace analysis that is capable of
        inferring the invariants under study from transaction history.

    \item \textbf{Experimental Results:} This paper presents the first
        systematic and quantitative evaluation of the effectiveness of runtime
        invariants on $42$ victim contracts in $27$ real-world exploits with high
        financial losses.
        

    \item \textbf{Invariant Study Findings:} Our research uncovers a series of
        critical insights that will inform the future application and
        development of dynamic invariants.
\end{itemize}


\vspace{-1ex}
\section{Background}
\vspace{-0.5ex}



\noindent\textbf{Blockchain} is a distributed and immutable ledger technology that records transactions across multiple nodes in a network. It employs cryptographic techniques to ensure data integrity and consensus algorithms to maintain final consistency across all participating nodes. \noindent\textbf{Smart contracts} are self-executing programs with the terms of the agreement directly written into code. Deployed on blockchains, they are immutable and transparent, enabling trustless transactions without the need for intermediaries. \noindent \textbf{Invariant guards \camera{(also called circuit breakers)}} are runtime checks around contract invariant conditions that shall always hold during contract execution, aiming to secure smart contracts on the fly.

\noindent\textbf{Ethereum Virtual Machine (EVM)}  is the runtime environment for smart contract execution on Ethereum. It is a Turing-complete virtual machine that interprets and executes the bytecode compiled from contracts programmed in a high-level language like Solidity. \noindent\textbf{Gas} is a unit of transaction fee on Ethereum, used to quantify the computational efforts for the execution of EVM operations. It is paid in \textbf{Ether}, the native cryptocurrency of Ethereum.
\noindent\textbf{Externally Owned Account (EOA) and contract account} are two types of accounts on Ethereum. EOAs are owned by normal users only who have the right to send transactions to blockchains, while contract accounts are controlled by the code deployed at a certain blockchain address and its code will be executed when the contract function is invoked. \noindent\textbf{ERC20} is a standard interface for fungible tokens on Ethereum. Almost all valuable tokens on Ethereum are ERC20 tokens.




\noindent \textbf{Common smart contract vulnerabilities} include integer overflow/underflow~\cite{BECoverflow},
reentrancy~\cite{DAOattacks}, dangerous delegatecall~\cite{Paritystolen}, etc.
\textbf{DeFi vulnerabilities }are smart contract vulnerabilities that are specific to DeFi applications. They are
more subtle to detect and attacks usually involve more sophisticated steps~\cite{zhou2023sok}.
DeFi protocols are major targets for smart contract attacks, which have experienced \$5.53 billion loss
out of the overall \$6.94 billion caused by recent blockchain incidents~\cite{defillama}.




%

\vspace{-1ex}
\section{Motivating Example}\label{sec:example}
\vspace{-1ex}

In this section, we present a motivating example to illustrate how an exploit
transaction behave differently from other benign transactions in histories and
how dynamically inferred transactions can neutralize the exploit transaction to
enhance smart contract security.

\noindent \textbf{Exploit Transaction:} Harvest Finance is a Decentralized
Finance (DeFi) protocol deployed on Ethereum to manage and auto-invest stable
coins for users. On October 26, 2020, USDC and USDT vaults of the Harvest
Finance were exploited, causing a financial loss of about USD \$$33.8$ million.
Harvest Finance internally uses the market data of Curve, another DeFi stable
coin trading protocol, to determine the market prices of USDC and USDT. In the
exploit transaction, the attacker distorts the market of Curve to cause the
Harvest Finance to make sub-optimal investment decisions.

Specifically, the attack transaction first borrows a large
amount of digital assets and uses the borrowed asset to buy USDC in Curve to
inflate its price. Then it deposits $49.98$M USDC into Harvest
vault contract, which increases its USDC balance from $72.83$M to $122.51$M.
Due to the manipulated oracle price of USDC, Harvest vault contract erroneously
mints the attacker an inflated $51.46$M fUSDC, which increases the total fUSDC
supply from $127.58$M to $179.04$M. The attacker then restores the USDC by
selling the USDC in Curve, and redeems all its fUSDC tokens for $50.30$M USDC,
yielding a $32$k surplus compared to the initial deposit. This redemption
decreases the Harvest vault's USDC balance from $122.81$M to $72.51$M and
restores the total fUSDC supply to its original value of $127.58$M.
Remarkably, this identical attack vector is executed three times
within one exploit transaction, consuming an unusually high gas count of
$9,895,111$, narrowly within the gas limit of $12,065,986$ at the time. 
\camera{More details of this example can be found at the website~\cite{website}.}

\begin{figure}[!htbp]
  \centering
  \begin{subfigure}[b]{0.98\textwidth}
      \centering
      \includegraphics[width=\textwidth]{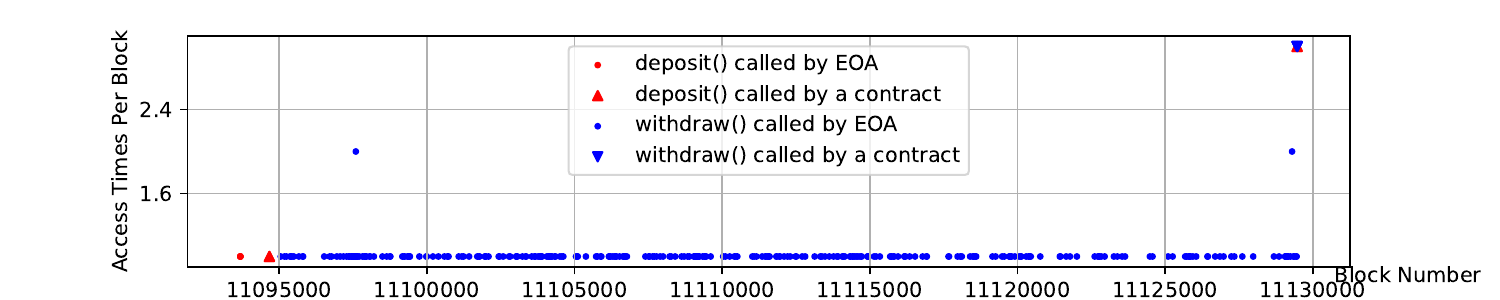}
  \end{subfigure}
  \hfill
  \begin{subfigure}[b]{0.98\textwidth}
      \centering
      \includegraphics[width=\textwidth]{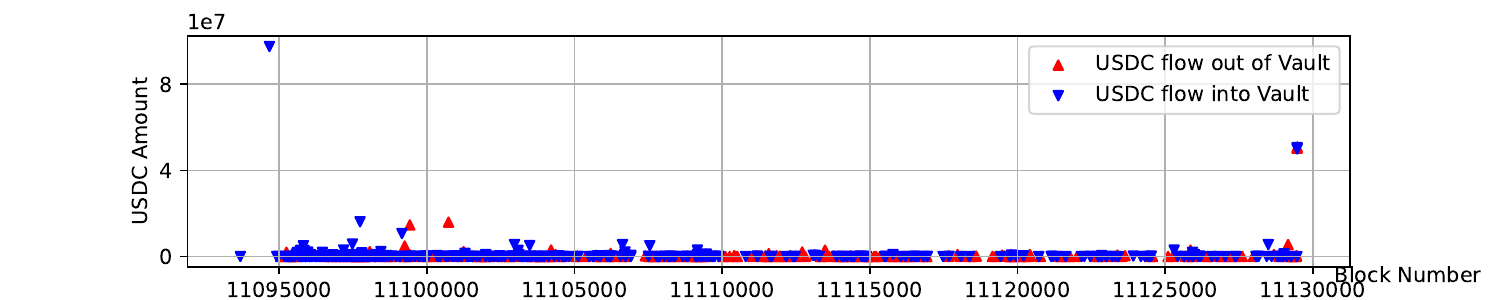}
  \end{subfigure}

  \begin{subfigure}[b]{0.98\textwidth}
      \centering
      \includegraphics[width=\textwidth]{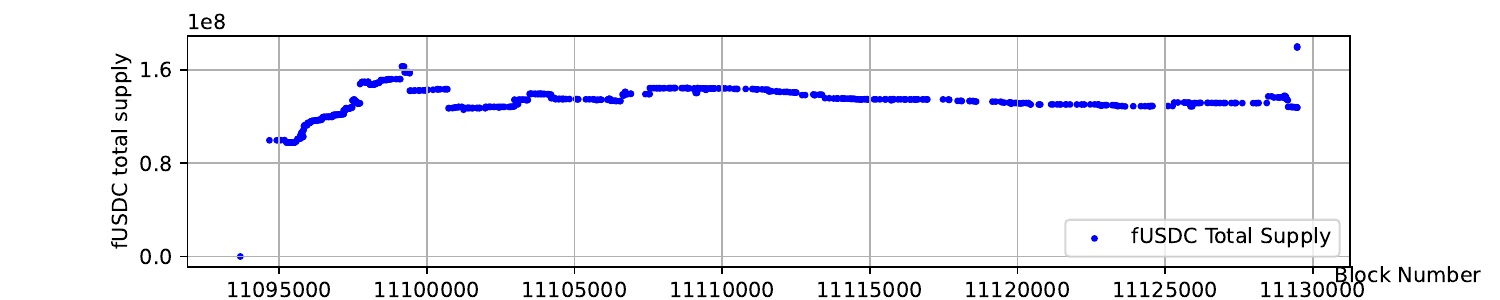}
  \end{subfigure}
  \hfill
  \begin{subfigure}[b]{0.98\textwidth}
      \centering
      \includegraphics[width=\textwidth]{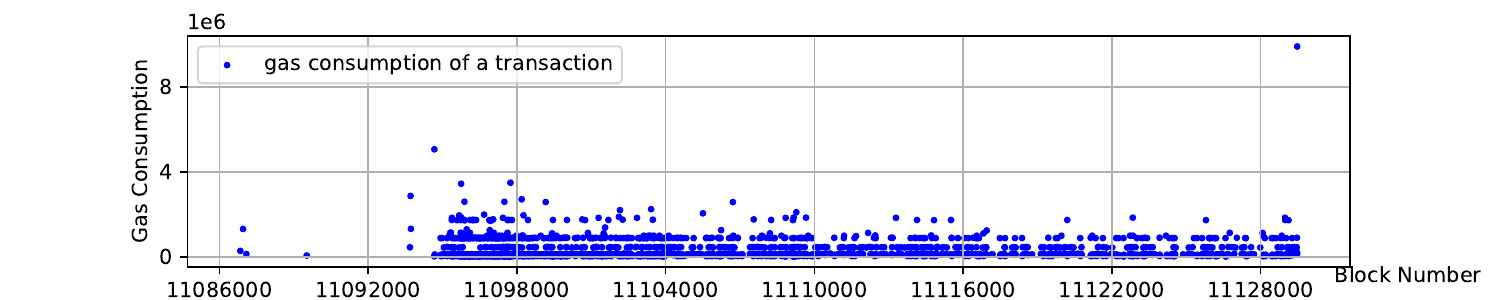}
  \end{subfigure}
  \caption{Statistics of Transactions on Harvest USDC Vault Contract.}
  \label{fig:harvest_abnormal}
\end{figure}

\noindent \textbf{Abnormal Behaviors:} We identify four distinct dimensions of
abnormal behavior: a high frequency of user interactions with the Harvest vault
contract, an exceptionally large volume of token flow, abrupt fluctuations in
the total supply of fUSDC tokens, and remarkably high gas consumption. To
better understand the abnormality of the exploit transaction, we collect and
analyze all transaction history of the Harvest vault contract up to the point
of the exploit, as illustrated in \Cref{fig:harvest_abnormal}. 

As shown in \Cref{fig:harvest_abnormal}, the last data point in each
sub-figure, representing the exploit transaction, consistently emerges as an
outlier. Specifically, we have observed that the exploit transaction is the
first in the contract's history to: (1) invoke the
\texttt{withdraw} function from a contract rather than from a user address, (2) call both \texttt{deposit} and
\texttt{withdraw} functions $3$ times within one transaction, (3)
consume more gas than any previous transaction, (4) withdraw more USDC from the
protocol than any other transaction, \revise{(5)} elevate the total supply of fUSDC
tokens to an all-time high. 

We observed several other outlier data points besides the last data point. 
For access frequency, two blocks contained two \texttt{withdraw} invocations within the same block, which, 
upon investigation, were found to originate from separate addresses. 
We believe it was a coincidence that two users withdrew their funds in the same block. 
For token flow, a very large USDC deposit occurred immediately after the contract's deployment, 
which, upon manual analysis, was identified as an initial liquidity provided by the Harvest Finance team. 
Similarly, the abrupt jump in the total supply of fUSDC tokens in the early stage was also due to this 
initial liquidity provision. 
Since users receive fUSDC tokens when they deposit USDC into the vault and these tokens are burned upon withdrawal of USDC,
a large initial deposit would lead to a corresponding increase in the total supply of fUSDC tokens.
For gas consumption, a transaction at block 11094671 consumed 5,062,934 gas, which, 
after manual analysis, was found to be sent by the Harvest Finance team for migrating old vaults to new ones. 
This migration process involved numerous operations in a single transaction to ensure atomicity, leading to high gas consumption. 
Overall, while there were anomalies during the early stage of the contract's deployment, 
the different transaction metrics stabilized after some blocks until significantly changed by the exploit transaction. 
If contract maintainers can identify the "stable" stage and use these transactions to infer invariants, 
the inferred invariants can be much more accurate. 
However, in this paper, we assume such information is not available.

\noindent \textbf{Apply Inferred Invariants:} The multi-dimensional
abnormalities observed in the exploit transaction highlight a stark departure
from typical transactional behaviors. This divergence suggests the feasibility
of crafting and applying runtime invariants that are capable of flagging and
blocking transactions exhibiting such anomalous characteristics. For example,
suppose we inferred and enforced an invariant stating that the
\texttt{withdraw} function may only be invoked by an Externally Owned Account (EOA),
the exploit transaction could be blocked. This is because the exploit relies on a
contract to execute complex logic designed to
extract funds from the vault. Likewise, if we inferred an invariant that the total
supply of fUSDC should not surpass $160$M, the profitability of each round of the exploit
would be significantly reduced, making it unable to cover the cost of manipulating
the market of Curve.
Importantly, both invariants do not affect any normal user's
transaction in histories, making them practical for real-world deployment.

\noindent
\revise{\noindent \textbf{Patch Smart Contracts Post-Deployment:} }
\revise{Despite the immutable nature of deployed smart contracts,
developers still have different methods to alter their behavior post-deployment,
allowing for the addition or modification of invariant guards to shield against future exploits:  
\textit{(1) Upgradable or Modular Contract Design:} Upgradable contract standards such as
ERC897~\cite{ERC897} and ERC1167~\cite{ERC1167} incorporate a proxy and an execution contract. The proxy contract delegates function calls to the execution contract,
whose address can be modified within the proxy, allowing developers to update their smart contracts after deployment. Similarly, 
developers can segment a protocol into multiple contracts as different modules. A primary contract interfaces with users, subsequently interact with other modular contracts that handle distinct functionalities including invariant checking. The addresses of modular contracts could be updated in the primary contract. 
\textit{(2) Application Interface Adjustments:} For deployed protocols with neither upgradable nor modular contract designs, addressing vulnerabilities or enforcing invariants can still be achieved by launching a revised protocol version and redirecting users through website or application interfaces. 
}

\noindent \textbf{Research Questions:} Inspired by these observations and their
implications for enhancing smart contract security, we are motivated to explore
the following research questions:

\textbf{RQ1: } Given the fact that exploit transactions often exhibit abnormal
behaviors, \textbf{what kinds of invariant guards are most effective at
stopping exploit transactions?}

\textbf{RQ2: } If an exploit transaction or benign transaction violates
invariant guards, \textbf{how difficult is it for an attacker or a regular user
to bypass them?}

\textbf{RQ3: } As multiple dimensions of abnormality may be associated with an
exploit transaction, \textbf{how effective is the combination of different
invariant guards in preventing exploits?}

\textbf{RQ4: } Invariant guards require additional gas at runtime. \textbf{What
are the gas overheads of different types of invariant guards?}

\revise{
    \textbf{RQ5: } In terms of enhancing smart contract security, 
    \textbf{how does this work compare to other state-of-the-art works in
     contract invariant generation or transaction anomaly detection?}
}

\vspace{-3mm}
\section{Invariants}\label{sec:invariants}
\vspace{-1mm}

\noindent
\textbf{Scope.}
Our research focuses on the invariants that can be used to distinguish between benign and malicious transactions.
Particularly, we focus on the invariants that are broadly applicable to most common DeFi protocols.
Invariants that are not for security purposes or highly specific to a single protocol are outside the scope of our study.

\noindent\textbf{Methodology.}
In our effort to build a comprehensive list of smart contract transaction
guards, we carried out an analysis of existing research papers on smart
contract security~\cite{liu2022finding, ghaleb2023achecker, rodler2018sereum, choi2021smartian, breidenbach2021chainlink}.

Additionally, we conducted a qualitative study on both audit reports and source code of the $63$
    audited projects from ConsenSys,
    a leading smart contract auditing firm, from May 2020 to March 2023~\cite{consensysAudits}.
    One author extracted $2,181$ enforced invariants from the audit reports and smart contracts under audit by
    searching for keywords such as ``require'' and ``assert'', after eliminating any duplicates.
    The author then manually reviewed these invariants to extract templates for pattern matching
    against remaining uncategorized invariants. This iterative process continued until no new
    templates could be extracted,
    and all remaining invariants were also deemed uncategorized for specific reasons. 
    This task took three weeks.
    Following this, another author reviewed and validated both the categorized
    and uncategorized invariants for accuracy.
    In cases of disagreement, a third author was consulted to
    resolve the issue. This review process lasted two weeks. 
    All three authors have over two years of smart contract security research experience.

\vspace{-1mm}
 \begin{table}[!htbp]
    \scriptsize
    \vspace{-1mm}
    \caption{\revise{Qualitative Study Statistics Overview$^+$ (The table's left section presents key statistics from the qualitative study.
    The middle section presents categorized instances across invariant categories.
    The right section presents instances for various reasons why these invariants remain uncategorized.)}}
    \vspace{-3mm}
    \label{tab:qualitative}
    \begin{tabular}{|ll|ll|ll|}
        \hline
    Statistics                  & Count & Category        & Count  & Reason                     & Count \\ \hline
    \# Audits                   & 63    & Access Control  & 283    & Protocol Specific          & 1098   \\
    \# Code Repositories        & 49    & Time Lock       & 158    & Array Length Check         & 200   \\
    \# Invariants in Total      & 2181  & Gas Control     & 2      & Byte Operation             & 44    \\
    \# Invariants Categoried    & 826   & Re-entrancy     & 12     & Safe Math                  & 13    \\
    \# Invariants Uncategorized & 1355  & Oracle Slippage & 15     &                            &       \\
                                &       & Special Storage & 24     &                            &       \\
                                &       & Money Flow      & 151    &                            &       \\
                                &       & Data Flow       & 181    &                            &       \\ \hline
    \end{tabular}
    \flushleft
    \scriptsize
    \camera{$^+$:All study results and collected invariant instances can be found at~\cite{study}}.
    \vspace{-3mm}
    \end{table}

    The above process resulted in $826$ invariants under $8$ categories, 
    which represent $37.87\%$ of the $2181$ invariants, as shown in~\Cref{tab:qualitative}.
    Access Control and Data Flow are the two common categories.
    The remaining invariants ($62.13\%$) were not categorized for various reasons.
    The most common reason is that invariants are specific to a particular protocol.
    For example, the invariant ``\textit{require(validUniswapPath(bAsset))}'' checks whether
    ``\textit{bAsset}'' is a valid Uniswap path.
    However, this invariant can only apply to protocols involving Uniswap, thus limiting its applicability.
    Other uncategorized invariants are used for checking array lengths, byte operations, and arithmetic 
    safety, which target specific data structures or operations. 
    Such invariants of low-level operations are hard to apply as security guards because they are unable 
    to capture the high-level user intentions.


\noindent\textbf{Invariant Templates.}
Table~\ref{tab:invariants} summarizes the results of our study. In the table,
the \emph{Category} column groups invariant templates based on their application
domains, such as Access Control, Time Lock, etc.
The \emph{ID} column assigns a unique identifier to each invariant, while the \emph{Name} column provides
a human-readable description.
The \emph{Template} column contains formal representation of the invariant
templates. Specifically, we use $\avar$ to denote a contract state record maintained by invariant templates.
We use $\areg$ to represent a local variable and $\_?$ as the undetermined
parameter to be inferred. The \emph{Parameter} column shows the type of the
undetermined parameter. The \emph{References} column lists the academic or
industry sources of each invariant template.


\begin{table}[!htbp]
    \scriptsize
    \vspace{-4mm}
    \caption{Invariants. (We use $\avar$ to denote a contract state variable, $\areg$ to denote a local variable, and 
    $\_?$ to denote a hole in the template to fill during the synthesis. $|\_|$ denotes the absolute value.) }
    \vspace{-4mm}
    \label{tab:invariants}
    \renewcommand*{\arraystretch}{1.2}
    \setlength{\tabcolsep}{4pt}
    \begin{tabular}{|l|l|l|l|l|l|}
        \hline
        Category                         & ID   & Name                    & Template                                                                & Parameter & References                                                                                                        \\ \hline
        \multirow{5}{*}{Access Control}  & EOA  & onlyEOA                 & $\msgsender = \txorigin$                                                & -         & \multirow{5}{*}{\begin{tabular}[c]{@{}l@{}}  \cite{liu2022finding} \\ \cite{ghaleb2023achecker} \end{tabular} }   \\ \cline{2-5}
                                         & SO   & isSenderOwner           & $\msgsender = \owner$                                                   & address   &                                                                                                                   \\ \cline{2-5}
                                         & SM   & isSenderManager         & $\msgsender =  \bigcup_{i=1}^{n?}\ \manageri $                          & addresses &                                                                                                                   \\ \cline{2-5}
                                         & OO   & isOriginOwner           & $\txorigin = \owner$                                                    & address   &                                                                                                                   \\ \cline{2-5}
                                         & OM   & isOriginManager         & $\txorigin =  \bigcup_{i=1}^{n?}\ \manageri$                            & addresses &                                                                                                                   \\ \hline
        \multirow{3}{*}{Time Lock}       & SB   & isSameSenderBlock       & $\flagentrySender \neq \flagexitSender $                                & -         & \multirow{2}{*}{\cite{idleFinance}}                                                                               \\ \cline{2-5}
                                         & OB   & isSameOriginBlock       & $\flagentryOrigin \neq \flagexitOrigin $                                & -         &                                                                                                                   \\ \cline{2-6} 
                                         & LU   & lastUpdate              & $\currblocktimstap - \lastblocktimstap \geq \nbBlocks $                 & Integer   & \cite{mstable, fei, feiAudit}                                                                                     \\ \hline
        \multirow{2}{*}{Gas Control}     & GS   & GasStartUpperBound      & $\gasStart \leq \gasUpperbound $                                        & Integer   & \multirow{2}{*}{motivated by \Cref{sec:example}}                                                                  \\ \cline{2-5}
                                         & GC   & GasConsumedUpperBound   & $\gasStart - \gasEnd \leq \gasUpperbound $                              & Integer   &                                                                                                                   \\ \hline
        Re-entrancy                      & RE   & nonReEntrant            & $\lockReentry = \mytrue $                                               & -         & \cite{rodler2018sereum}                                                                                           \\ \hline
        \multirow{2}{*}{Oracle Slippage} & OR   & OracleRange             & $\priceLowerbound \leq \priceflag \leq \priceUpperbound $               & Integer   & \cite{dforce2}                                                                                                    \\ \cline{2-6} 
                                         & OD   & OracleDeviation         & $| (\priceflag - \oldpriceflag) / \oldpriceflag | \leq \priceDeviation$ & Integer   & \begin{tabular}[c]{@{}l@{}}  \cite{dforce2} \\ \cite{breidenbach2021chainlink} \end{tabular}                      \\ \hline
        \multirow{2}{*}{Special Storage} & TSU  & TotalSupplyUpperBound   & $\flagtotalSupply \leq \totalSupplyUpperbound $                         & Integer   & \cite{AaveV3} \cite{dforce3}                                                                                      \\ \cline{2-6} 
                                         & TBU  & TotalBorrowUpperBound   & $\flagtotalBorrow \leq \totalBorrowUpperbound$                          & Integer   & \cite{AaveV3} \cite{dforce3}                                                                                      \\ \hline
        \multirow{4}{*}{Money Flow}      & TIU  & TokenInUpperBound       & $\tokenInCap \leq \valueVar$                                            & Integer   & \cite{balancer} \cite{dforce3}                                                                                    \\ \cline{2-6} 
                                         & TIRU & TokenInRatioUpperBound  & $\tokenInCap \leq \valueVar$                                            & Integer   & \cite{balancer}                                                                                                   \\ \cline{2-6} 
                                         & TOU  & TokenOutUpperBound      & $\tokenOutCap / \tokenBalance \leq \valueVar$                           & Integer   & \cite{balancer} \cite{dforce3}                                                                                    \\ \cline{2-6} 
                                         & TORU & TokenOutRatioUpperBound & $\tokenOutCap / \tokenBalance \leq \valueVar$                           & Integer   & \cite{balancer}                                                                                                   \\ \hline
        \multirow{4}{*}{Data Flow}       & MU   & MappingUpperBound       & $ \mapIndexed{\indexMap} \leq \valueVar$                                & Integer   & \multirow{4}{*}{\cite{choi2021smartian}}                                                                          \\ \cline{2-5}
                                         & CVU  & CallValueUpperBound     & $\msgvalue \leq \msgV$                                                  & Integer   &                                                                                                                   \\ \cline{2-5}
                                         & DFU  & DataFlowUpperBound      & $\var \leq \valueVar$                                                   & Integer   &                                                                                                                   \\ \cline{2-5}
                                         & DFL  & DataFlowLowerBound      & $\var \geq  \valueVar$                                                  & Integer   &                                                                                                                   \\ \hline

        \end{tabular}
\end{table}

\vspace{-2ex}
\subsection{Access Control \camera{(also called Permission Control)}}
\vspace{-0.5ex}
Many research papers have conducted extensive studies on the access control~\cite{liu2022finding, ghaleb2023achecker}. Access control
governs the privileges associated with the transaction's sender and origin,
dictating which addresses are authorized to invoke specific smart contract
functions.

Note that transaction's sender and origin could be different in Ethereum. The sender
is the address which invokes the contract function, while the origin is the
address who initiates the entire transaction. For example, if user address $a$
calls contract $b$ which in turn calls contract $c$, during the execution of
$c$, the sender address is $b$ while the origin address is $a$.

\noindent\textbf{\emph{onlyEOA (EOA)}}\footnote{\camera{It is important to note that the forthcoming Spectra upgrade of Ethereum, anticipated for late 2024 or early 2025, will implement EIP-3074~\cite{EIP3074}. This implementation is expected to make EOA completely bypassable.}} This template verifies that the transaction's
origin matches the sender's address, thereby confirming it was initiated from
an externaly owned user address (i.e., EOA address) rather than a
contract address. The intuition of this invariant template is that many attack
strategies involves multiple sophisticated interactions and therefore attackers
often have to write their own contracts. This template can neutralize such
attack strategies.

\noindent\textbf{\emph{isSenderOwner (SO)}} and \textbf{\emph{isOriginOwner (OO)}}: These
templates restrict function execution to a predefined address ($\owner$) that
are registered as owners.

\noindent\textbf{\emph{isSenderManager (SM)}} and \textbf{\emph{{isOriginManager (OM)}}}: These
templates only allow function calls from a set of predefined manager addresses
$\manageri$.

The access control invariants are typically inserted at the
beginning of non-read-only functions to immediately halt unauthorized attempts
to alter contract state.

\vspace{-2ex}
\subsection{Time Lock}
\vspace{-0.5ex}

The Time Lock category of invariants serves as a temporal gating mechanism for
smart contract functions. This category contains three invariants.

\noindent\textbf{\emph{isSameSenderBlock (SB)}} and \textbf{\emph{isSameOriginBlock (OB)}}: These templates
limit the ability to execute specific paired functions within the same block by
the same sender or origin. For example, to inhibit the same sender or origin
address from invoking both the deposit and withdraw functions consecutively
within a single block. The intuition is that normal users are unlikely to
initiate multiple interactions with the same function in a few seconds, while
malicious attackers often use iterative loops to drain funds from a victim
contract. To implement these invariants, a state variable $\flagentrySender$ (resp.,
$\flagentryOrigin$) stores a hashed combination of the transaction sender
(resp., origin) address and the current block number upon entry into a function
(e.g., deposit). In the exit function (e.g., withdraw), this stored value is
compared against a freshly computed hash, stored in $\flagexitSender$ (resp.,
$\flagexitOrigin$), to ensure that they differ. These two invariants are
designed to be updated at the entry point of \emph{enter} functions, i.e.,
functions that accept tokens from users. Then verified at the start of
\emph{exit} functions, i.e., functions that are responsible for disbursing
tokens back to users.

\noindent\textbf{\emph{lastUpdate (LU)}}: These template moderates the frequency with
which a given function can be invoked. It inserts guard at the beginning of
non-read-only functions to mandate that a specified number of blocks, denoted
as $\nbBlocks$, must elapse between two consecutive calls to the same function.
To enforce this, the state variable $\lastblocktimstap$ captures the timestamp
of the last block where the function was invoked. Subsequent calls to the
function check this stored timestamp against the current block timestamp.
The difference must meet or exceed the $\nbBlocks$ threshold.

\vspace{-2ex}
\subsection{Re-entrancy}
\vspace{-0.5ex}
The Re-Entrancy class of invariants tackles re-entrancy vulnerabilities in
smart contracts. Represented by a single invariant template, \textbf{\emph{nonReEntrant
(RE)}}, this category utilizes a state variable $\lockReentry$ as a lock to prevent
a transaction from entering a set of key functions of a contract more than once.
$\lockReentry$ will be set to $True$ when a function is invoked and reset to
$False$ when a function returns. The RE guard is usually placed at the
beginning of \emph{enter} and \emph{exit} functions of a contract to effectively mitigate
re-entrancy risks.

\vspace{-2ex}
\subsection{Gas Control}
\vspace{-0.5ex}
We propose the Gas Control category of invariants, motivated by the
\emph{Harvest} example. The intuition is that malicious attacks tend to have
significantly more complicated logic to consume a large amount of gas. This
class consists of two invariants: \textbf{\emph{GasStartUpperBound (GS)}} and
\textbf{\emph{GasConsumedUpperBound (GC)}}. As illustrated in
Table~\ref{tab:invariants}, the GS invariant sets an upper limit on the
remaining gas at the entry point of a function, using the variable $\gasStart$,
whereas the GC invariant sets an upper bound on the total gas consumed
within the function by comparing the remaining gas at the entry and exit point
of a function, $\gasStart$ and $\gasEnd$. These invariants are designed to be placed at the beginning and end of
non-read-only functions.

\vspace{-2ex}
\subsection{Oracle Slippage}
\vspace{-0.5ex}
The Oracle Slippage category mitigates risks tied to price oracles in DeFi applications. The intuition of templates in this
category is to detect potential price manipulation by malicious attacks.
This class includes two invariant templates: \textbf{\emph{OracleRange (OR)}} and
\textbf{\emph{OracleDeviation (OD)}}. The OR template enforces a bounded range
for the oracle prices. It utilizes two parameters, $\priceLowerbound$ and
$\priceUpperbound$. The OD template enforces a specific percentage
deviation limit between the current and last price provided by the oracle. The
parameter $\priceDeviation$ is employed to define a
permissible deviation rate.
These invariants are usually inserted right after the oracle is called.

\vspace{-2ex}
\subsection{Special Storage}
\vspace{-0.5ex}
The Special Storage class of invariant templates is concerned with constraining
global storage variables that are crucial to the contract's state or logic. The intuition
of this is after an exploit, the contract's state variables are often in an abnormal state.
Thus, by constraining the state variables, we can prevent the exploit. This class features two main invariants: \textbf{\emph{TotalSupplyUpperBound (TSU)}} and
\textbf{\emph{TotalBorrowUpperBound (TBU)}}. The TSU invariant imposes an upper
bound (denoted as $\totalSupplyUpperbound$) on the contract's totalSupply variable,
while TBU sets a ceiling (denoted as $\totalBorrowUpperbound$) for the contract's
totalBorrow variable which represents the total amount that can be borrowed from the contract.
To preserve the integrity
of these important state variables, these invariants are inserted at the functions that could modify the total supply or total
borrow balances.

\vspace{-2ex}
\subsection{Money Flow(also called Token Flow)}
\vspace{-0.5ex}
The Money Flow class focuses on the flow of tokens within the smart contract,
particularly for functions involving token deposits and withdrawals. The
intuition of these templates is that malicious transactions tend to cause
abnormally large \revise{amount} of digital asset movement.

\noindent\textbf{\emph{TokenInUpperBound (TIU)}} and \textbf{\emph{TokenOutUpperBound (TOU)}}:
This template caps the number of tokens flowing into or out of the contract
each time by using an integer parameter $\valueVar$.

\noindent\textbf{\emph{TokenInRatioUpperBound (TIRU)}} and \textbf{\emph{TokenOutRatioUpperBound
(TORU)}}: These templates constrain the ratio of tokens flowing into or out of
the contract, in relation to the contract's current token balance. They also
employ an integer parameter $\valueVar$.

To ensure effective governance of money flow, these invariants are placed
within functions that handle token transfers. The TIU and TIRU
invariants are applied right before a token deposit, while the TOU and
TORU invariants are applied before a token withdrawal.

%
%
%

\vspace{-2ex}
\subsection{Data Flow}
\vspace{-0.5ex}
Smartian~\cite{choi2021smartian} leverages dynamic taint analysis to detect whether
a block state can affect an ether transfer. We extend their work to include all data flows affecting both ether and ERC20 token transfers.
This allows us to set constraints on values that could potentially be controlled by an attacker to manipulate transfer amounts. The Data Flow category is subdivided into four specific invariants, each designed to address a particular type of variables in the data flow.

\noindent\textbf{\emph{{MappingUpperBound (MU)}}}: This invariant focuses on values stored in the contract's mapping data structure, often representing a user's property(e.g., a user's shares/balances). To constrain such user-specific values, we introduce a parameter $\valueVar$ to set an upper limit on these mappings.

\noindent\textbf{\emph{{CallValueUpperBound (CVU)}}}: Call values signify the amount of ether transferred during a function call. Because these values directly affect the contract's ether balance, we list it as a separate invariant. We employ a parameter, denoted as $\msgV$, to cap the incoming ether to mitigate risks of abnormal or malicious deposits.

\noindent\textbf{\emph{DataFlowUpperBound (DFU)}} and \textbf{\emph{DataFlowLowerBound (DFL)}}: These invariants apply to all other data flow variables, whether derived from external calls, storage loads, or calldata. To regulate these variables, we use a parameter $\valueVar$, setting either upper or lower bounds on these values to thwart unauthorized manipulations.

The above invariants are inserted at the locations where data flow variables
are first read. This is often in functions that initiate token transfers.
These invariants help the contract ensure that every value used for the calculation of token transfers is within the normal range.

\section{{\name}}

\begin{figure}[!htbp]
  \centering
  \includegraphics[width=1.0\textwidth]{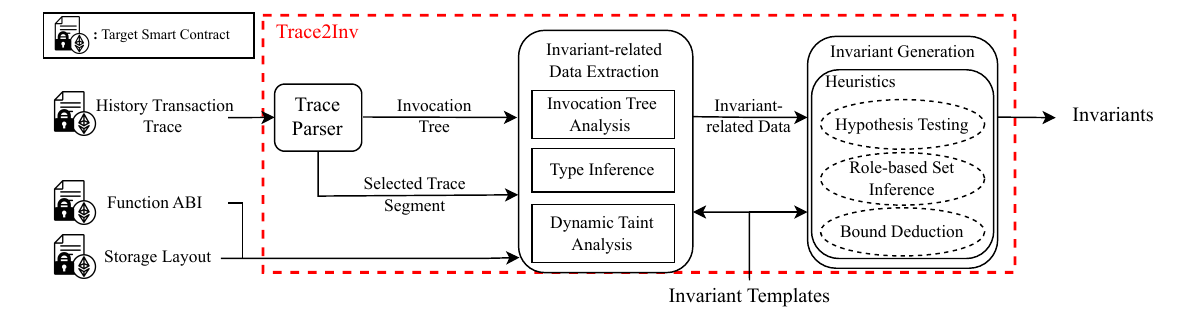}
  \vspace{-9mm}
  \caption{An Overview of {\name}}
  \label{fig:overview}
  \vspace{-6mm}
\end{figure}

We present {\name}, shown in \Cref{fig:overview}, a framework to infer concrete invariants as introduced in Section~\ref{sec:invariants} for a contract by analyzing its transaction traces. {\name} consists of three modules: trace parser, invariant-related data extraction, and invariant generation. The second 
module consists of three submodules: invocation tree analysis, type inference, and dynamic taint analysis.

\subsection{Trace Parser}

\revise{A transaction's trace data, denoted as \emph{structLogs}}, contains a sequence of executed EVM instructions and each instruction is a six-item tuple $\mathsf{\langle pc, op, gasLeft, gasCost, stack, memory \rangle}$ that includes the current program counter, the EVM opcode to execute, the amount of the remaining gas and the gas consumed by current opcode execution, the full view of current EVM stack and memory.
In trace parser, we reconstruct the functional context information from \emph{structLogs} as an invocation tree where each node represents an external function call and the node hierarchy reflects the call-chain relationship. In particular, each tree node is a seven-item tuple $\mathsf{\langle addr, func, args, ret\_data, ins, gasEntry, gasExit \rangle}$ that records the current contract address, function name, corresponding arguments, returned data, the set of executed EVM instructions belonging to the function, and the amount of the remaining gas at the entry and exit points of the function. \revise{The invocation tree also contains metadata of the transaction, such as the transaction hash, the block number, and the origin address.
The trace parser leverages the target contract address to isolate a selected trace segment corresponding to the target contract's execution. This invocation tree and the segmented trace data are then passed to the next module for further analysis.}

\vspace{-2ex}
\subsection{Invocation Tree Analysis}
\vspace{-1ex}
Invocation tree analysis extracts invariant-related data from the invocation tree,
that is sufficient to collect data for \emph{access control},
\emph{time lock}, \emph{gas control}, \emph{re-entrancy}, \emph{oracle}, and \emph{money flow} invariants.
For instance, for the \emph{sender} opcode, its results is extracted 
from the invocation tree by searching for parent node of 
the target contract node. 
Moreover, the invocation tree is also used to identify 
locations of \emph{re-entrancy} by capturing nested 
and recursive calls to the target contract. 
Oracle values are also obtained by traversing the 
invocation tree to locate calls to the oracle. 
For money flow invariants, the required values are also read 
from function calls of transferring Ether/ERC20.

\vspace{-2ex}
\subsection{Type Inference}
\vspace{-1ex}
Type inference in {\name} infers the type of storage slot when it is accessed. It is essential for extracting data relevant to both \emph{special storage} and \emph{data flow} invariants, as accurately decoding storage accesses with the correct type is necessary for those invariants. Decoding storage slots is straightforward when they are listed in the contract's storage layout, which is typically the case for \emph{special storage} invariants.
However, the challenge arises with complex data structures like \emph{mapping}, often used in \emph{data flow} invariants. These structures may reference storage slots not explicitly present in the contract's storage layout, which are instead computed through operations such as \emph{sha3} and arithmetic functions.
To solve this issue, {\name} maintains a preimage dictionary to track the key's computation. 
When we encounter a \emph{sha3} opcode, the mapping positions and slots are recorded. In Solidity, the first $32$-bytes represent the mapping slot, and the next $32$-bytes serve as the mapping position. In Vyper, these roles are reversed. Anytime a 64-byte sha3 hash is encountered, both its hash value and origin are recorded.
When a \emph{sload} operation has a key that is not present in the storage layout, the preimage dictionary is consulted. We recursively trace back the computation steps until we identify a mapping data structure that is in the storage layout. Using the types of the mapping data structure, we infer the type of storage slot accessed.


\subsection{EVM-level Dynamic Taint Analysis}
\revise{Dynamic taint analysis in {\name}
operates at the EVM opcode level to track data sources. It is used to extract data for \emph{data flow} invariants.}
Given a target contract, a corresponding trace segment, and the invocation tree of the transaction, the taint analyzer collects all accessible information pertaining to that contract. 
The analyzer also infers the data types of recorded tainted data or taint sources. It accomplishes this by utilizing storage layouts and function ABIs thereby providing a comprehensive, type-aware taint analysis tailored for smart contracts. 


\begin{table}[!htbp]
    \caption{Instructions Defined as Sources and Sinks.}
    \vspace{-4mm}
    \label{tab:taintsSinks}
    \scriptsize
    \begin{tabular}{|l|l|l|}
        \hline
                                 & Category                    & Opcodes and Locations                                                                              \\ \hline
        \multirow{6}{*}{Sources} & External Address            & balance, extcodesize, extcodecopy, extcodehash                                                     \\ \cline{2-3} 
                                 & Execution Context           & origin, caller, address, codesize, selfbalance, pc, msize, gas                                     \\ \cline{2-3} 
                                 & Call Data                   & callvalue, calldataload, calldatasize, calldatacopy                                                \\ \cline{2-3} 
                                 & Return Data                 & returndatasize, returndatacopy                                                                     \\ \cline{2-3} 
                                 & Block                       & blockhash, coinbase, timestamp, number, prevrandao, gasprice, gaslimit, chainid                    \\ \cline{2-3} 
                                 & Storage                     & sload(untainted)                                                                                   \\ \hline
        \multirow{4}{*}{Sinks}   & ether transfer              & address.call\{value: \textbf{uint256}\}                                                            \\ \cline{2-3} 
                                 & ether transferFrom          & callvalue                                                                                          \\ \cline{2-3} 
                                 & ERC20 transfer              & ERC20.transfer(address,\textbf{uint256}), ERC20 = token                                            \\ \cline{2-3} 
                                 & ERC20 transferFrom          & ERC20.transferFrom(address1,address2,\textbf{uint256}), ERC20 = token, address2 = this             \\ \hline
        \end{tabular}
        \vspace{-3mm}
    \end{table}
    

\noindent\textbf{Taint Sources and Sinks.} Table \ref{tab:taintsSinks} lists the EVM instructions that our dynamic taint analyzer identifies as sources and sinks for taint propagation. The table is divided into two categories: \emph{Sources} and \emph{Sinks}. In the \emph{Sources} category, we outline various sub-categories of taint sources, which include external address, execution context, call and return data, block variables, and storage. Opcodes like \emph{balance}, \emph{extcodesize}, and \emph{sload} are some of EVM opcodes that load new taint sources into the stack or memory. They are key to the taint propagation as they introduce data that could potentially influence other data points or outcomes in the contract.

On the other hand, the \emph{Sinks} category highlights areas where tainted data may potentially lead to undesired or vulnerable behaviors. These include operations like ether transfers and ERC20 token transfers. Notably, the locations of these sinks are marked in bold text, such as $\mathsf{address.call\{value: \textbf{uint256}\}}$ and $\mathsf{transfer(address,\textbf{uint256})}$, emphasizing their critical role in the taint analysis. Through these identified sources and sinks, our taint analyzer can effectively trace the flow of sensitive information within the smart contract, providing a robust framework for dynamic invariant inference and validation.

\noindent\textbf{Bit Level Taint Propagation.} Our taint analyzer maintains three distinct taint trackers: the stack, the memory, and the storage trackers. Initially, all these trackers are empty. Each tracker records taint information at the bit level, enabling granular analysis. For instance, \emph{sload} and \emph{sstore} opcodes can only read and write $32$-byte chunks, but the taint status is stored for each individual bit. Our propagation of taints follows a set of rules based on prior work~\cite{ghaleb2022etainter}:

\noindent\textbf{R1:} A value derived from one or more operands becomes tainted if any of the operands is tainted. 

\noindent\textbf{R2:} For \emph{sload}, if the storage slot loaded is tainted, the $32$-byte stack entry receives the corresponding taint information. If not, the \emph{sload} acts as a new taint source and taints the stack entry. For \emph{sstore}, the $32$-byte entry in the storage tracker is overwritten with the taint information from the stack.

\noindent\textbf{R3:} For instructions that read data from memory (e.g., \emph{mload}), the result value is tainted if the read data is tainted. The same logic applies for data loaded from storage using the \emph{sload} opcode. 

\noindent\textbf{R4:} In external calls, if the arguments in memory are tainted, the return data will also be tainted.  



\subsection{\revise{Invariant Generation}}
\label{subsec:inference}

Using the extracted data by the previous module, the invariants' inference module uses the templates provided in Table~\ref{tab:invariants} to generate concrete invariants where all holes are filled with concrete values.
The synthesis heuristic of generating invariants varies based on their category: 

\noindent\textbf{\revise{Hypothesis Testing}:} For EOA, SB, OB, and RE invariants, which act as assumptions regarding the contract's behaviors, no parameters need to be learned. If no data points violate these invariants (i.e, all transactions in the training set satisfy the invariant), they are then directly applied to the contract. Otherwise, the violated invariant is not applied. 

\noindent\textbf{\revise{Role-based Set Inference}:} For address-based invariants such as SO, SM, OO, and OM, we adopt a set-based heuristic. We analyze the set of senders or origins associated with each function call. If the size of this set exceeds a certain threshold (greater than $1$ for owners or $5$ for managers), we consider it as a violation, and the corresponding invariant is not applied to that function.

\noindent\textbf{\revise{Bound Deduction}:} Invariants in the categories of gas control, oracle slippage, special storage, and money flow require learning the bounds of certain integer parameters. For those, the maximum and minimum values are chosen among the collected data points, provided the data contains at least two distinct values. Particularly for the oracle slippage invariant (OR), an additional tolerance of $20$\% is included for the upper and lower bounds, in accordance with prior research~\cite{tolmach2021formal}. For TIRU and TORU invariants, we filter outliers using a \emph{z-score} threshold of $3$.

\noindent\textbf{\revise{Hybrid}:} Lastly, for the LU invariant, which deals with the block gap between calls to the same function, we calculate the smallest block gap among the data points. If any two data points have a block gap of \emph{zero}, the invariant is not applied to the corresponding function. Otherwise, the smallest block gap is used as the parameter for the invariant.

\subsection{{\name} Implementation}
\revise{
  {\name}'s input, transaction trace data, can be obtained from Ethereum via its API \emph{debug\_trace-Transaction}. 
  Within the Trace Parser module,
  {\name} queries EtherScan~\cite{Etherscan} to gather additional transaction metadata not directly available in the trace data, such as block number and transaction origin.  This metadata provides the execution context crucial for generating certain invariants, like EOA and LU.
  Additionally, {\name} also queries EtherScan~\cite{Etherscan} to fetch the contract's source code. The code is then compiled locally using the appropriate versions of the Solidity~\cite{solidity} or Vyper~\cite{vyper} compilers to obtain the contract's function ABI. With this ABI, {\name} leverages Slither~\cite{slither} to decode function names, their arguments, and return values present in the trace.  
  Within Invariant Related Data Extraction Module, {\name} compiles source code of target smart contract, and reads the compiler's output 
  to obtain its storage layout. However, some old versions of Solidity and Vyper compilers do not support this functionality. In such cases, {\name} fetches the storage layout from EVM Storage~\cite{EVMstorage}.  }

Several optimizations are incorporated into {\name} to enhance its performance. First, when fetching trace data from an Ethereum archive node, we optimize the process by using \textbf{batching RPC requests}. This significantly reduces the overhead associated with individual API calls. Second, we use \textbf{parallelization} techniques to speed up the fetching and parsing trace data. Specifically, multiprocessing is used to concurrently handle different segments of trace data, converting them into summaries in a time-efficient manner. Third, \textbf{caching} is utilized to further optimize performance; the system caches query results from EtherScan, archive nodes, and compilers. This minimizes redundant executions and API calls, thereby accelerating the overall analysis process. 



\vspace{-2ex}
\section{Evaluation}
\label{sec:evaluation}
In this section, we aim to empirically answer the research questions raised in~\Cref{sec:example} 
by applying invariant guards to real-world exploits on Ethereum Blockchain. 
We systematically collected economic exploit incidents that cost greater than 
300K USD financial loss from February 14, 2020, to \revise{August 1, 2022} on Ethereum Blockchain.
Our benchmark is compiled from a diverse set of sources including academic
publications~\cite{chen2022flashsyn, qin2021attacking}, industry databases~\cite{Hacked,
BlockSecMedium}, and open-source GitHub repositories~\cite{DeFiHackLabs,LearnEVMHacks}. It is worth
noting that we exclude from our benchmarks any hacks targeting individual user wallets, as these
are primarily the result of private key leakage, rather than protocol vulnerabilities. We also
exclude hacks where the victim contracts are close-source, as our manual analysis requires the
source code of the victim contracts.
Note that \name can also be applied to close-source contracts, as long as their function ABIs and
storage layout are available.

\section{Benchmarks}

\subsection{Criteria}

We systematically collect benchmarks from different resources. 

Our criteria:

1. Historic smart contract attacks on Ethereum between 02/14/2020 and 06/16/2022

2. With a financial loss > 300K USD. Victim protocols' derivative tokens are not counted because these tokens became worthless after the attack.

3. The financial loss is from the victim protocol. 

A counterexample is, an attacker exploited a vulnerability in a protocol A, use this vulnerability to mint lots of token ERC20(A), finally exchange ERC20(A) for some stable coins from Uniswap. In this case, even though the vulnerability is in protocol A, the financial loss is from Uniswap. We do not consider this case.

Another counterexample is, an attacker gained access to a smart contract, use this privilege to steal some tokens from users who approved this smart contract, stealing money from users. 
In this case, even though the vulnerability is in the smart contract, the financial loss is from users. We do not consider this case.

\Cref{tab:benchmarks} presents the benchmark dataset in our study. It comprises 27 hacks which resulted in financial losses over $2$ billion USD. The column \emph{FL} denotes whether the exploit involves flash loans, which require atomicity for the hack transactions. The column \emph{Type} denotes the type of victim contracts.
For each hack, two types of victim contracts are manually identified: payload (P) refers to the
 contract that eventually transfers abnormal amounts of tokens out of the protocol, and
interface (I) refers to the contract that is directly invoked by users to initiate this 
transfer. The \emph{History} column gives the length of the transaction history up to the
hack transaction for each victim contract.

\subsection{RQ1: Effectiveness of Smart Contract Invariants}
\label{sec:rq1}

\textbf{Experiment.} In our first experiment, we utilize $23$ pre-defined invariant templates,
as detailed in \Cref{sec:invariants}, to dynamically infer invariants from the transaction histories
of $42$ different victim contracts, using the invariant generation methods as described in
\Cref{subsec:inference}. We divide each contract's transaction history into two distinct sets: 70\% of
the transactions are allocated for training set, while the remaining 30\% are used as the test set.

To validate the effectiveness of the invariants dynamically inferred, we employ the transaction
trace data in the test set for evaluation. Utilizing the same parser and dynamic taint analyzer, we
obtain the invocation tree and data points pertinent to the particular invariant for each
transaction. With these, we can evaluate whether a transaction violates any of the invariant guards
in place. If a transaction is blocked by these invariant guards, it serves as a positive example
for the effectiveness of the invariants.
The validation process discriminates between different kinds of positives. Specifically, if the exploit transaction is successfully blocked by the invariant, it is categorized as a \emph{True Positive}. On the other hand, any non-exploit transactions blocked are counted as \emph{False Positives}.






\newpage

\vspace{-3mm}
\begin{table}[h]
    \caption{Complete Results of Invariant Effectiveness Evaluation. 
    (\colorbox[HTML]{D9D9D9}{Gray} cells indicate the exploit transaction is blocked 
    by corresponding invariant guard(True Positives). 
    The number in the cell represents the number of non-exploit transactions blocked 
    by corresponding invariant guard(False Positives). 
    Symbol \textbf{-} represents that the invariant is by nature not applicable to the benchmark contract. 
    Symbol \textbf{\xmark} represents that the transactions in the training set violate the invariant. 
    Symbol \textbf{$\emptyset$} represents the transactions in the training set do not have enough data points 
    to learn the invariant.)}
    
    \label{tab:rq1}
    \begin{adjustbox}{angle=90}
        \scriptsize
        \tabcolsep=0.11cm

        \begin{tabular}{l|l|lllll|lll|ll|l|ll|ll|llll|llll}
                                                    &                            & \multicolumn{5}{c}{Access Control}                                                                                                                      & \multicolumn{3}{c}{Time Lock}                                                              & \multicolumn{2}{c}{Gas Control}                              & \multicolumn{1}{c}{Re}     & \multicolumn{2}{c}{Oracle}                                    & \multicolumn{2}{c}{Storage}                                    & \multicolumn{4}{c}{Money Flow}                                                                                                & \multicolumn{4}{c}{Data Flow}                                                                                               \\ 
\rot{Kind}                                          & Benchmarks                 & \rot{EOA}                    & \rot{SO}                     & \rot{SM}                   & \rot{OO}                     & \rot{OM}                      & \rot{SB}                     & \rot{OB}                     & \rot{LU}                     & \rot{GS}                      & \rot{GC}                     & \rot{RE}                   & \rot{OR}                      & \rot{OD}                      & \rot{TSU}                     & \rot{TBU}                      & \rot{TIU}                    & \rot{TOU}                    & \rot{TIRU}                   & \rot{TORU}                     & \rot{MU}                     & \rot{CVU}                    & \rot{DFU}                    & \rot{DFL}                     \\  \toprule
                                                    & RoninNetwork               & 0                            & \xmark                       & \xmark                       & \xmark                     & \xmark                        & \xmark                     & \xmark                     & \xmark                       & \cellcolor[HTML]{D9D9D9}0     & 0                            & 0                          & -                             & -                             & -                             & -                             & 0                            & \cellcolor[HTML]{D9D9D9}0    & 0.4                          & \cellcolor[HTML]{D9D9D9}0.1   & -                            & 0                            & \cellcolor[HTML]{D9D9D9}0    & 0                             \\
                                                    & HarmonyBridge              & 0                            & 0                            & \xmark                       & \xmark                     & 0                             & 0                          & 0                          & \xmark                       & 2.8                           & 2.8                          & 0                          & -                             & -                             & -                             & -                             & 0                            & \cellcolor[HTML]{D9D9D9}0    & 0                            & \cellcolor[HTML]{D9D9D9}0     & -                            & 0                            & \cellcolor[HTML]{D9D9D9}0    & 0                             \\
                                                    & Nomad                      & 0                            & \xmark                       & 0                            & \xmark                     & \xmark                        & 0                          & 0                          & \xmark                       & 0                             & 0                            & 0                          & -                             & -                             & -                             & -                             & 0                            & 0                            & 0                            & \cellcolor[HTML]{D9D9D9}0     & -                            & -                            & 0                            & 0                             \\
\multirow{-4}{*}{\vertical{Bridges}}                & PolyNetwork                & 0                            & 0                            & \xmark                       & 0                          & \xmark                        & 0                          & \xmark                     & 0                            & 0                             & 0                            & 0                          & -                             & -                             & -                             & -                             & 0                            & 0                            & 0                            & 0                             & -                            & 0                            & -                            & -                             \\\midrule     
                                                    & bZx2                       & \cellcolor[HTML]{D9D9D9}0    & 0                            & 2.3                          & 0                          & 2.3                           & 0                          & 0                          & 3.3                          & 1.4                           & 1.9                          & 0                          & $\emptyset$                   & $\emptyset$                   & 0.5                           & 0                             & \cellcolor[HTML]{D9D9D9}0    & 0                            & -                            & -                             & 0                            & -                            & \cellcolor[HTML]{D9D9D9}0    & \cellcolor[HTML]{D9D9D9}0.5   \\
                                                    & Warp                       & 0                            & 0                            & \xmark                       & 0                          & \cellcolor[HTML]{D9D9D9}2.3   & 0                          & 0                          & \cellcolor[HTML]{D9D9D9}0    & \cellcolor[HTML]{D9D9D9}6.8   & \cellcolor[HTML]{D9D9D9}6.8  & 0                          & -                             & -                             & -                             & \cellcolor[HTML]{D9D9D9}93.2  & 0                            & \cellcolor[HTML]{D9D9D9}4.5  & 0                            & \cellcolor[HTML]{D9D9D9}0     & -                            & -                            & \cellcolor[HTML]{D9D9D9}4.5  & 0                             \\
                                                    & Warp\_I                    & \cellcolor[HTML]{D9D9D9}0    & 0                            & 11.1                         & 0                          & 33.3                          & -                          & -                          & 0                            & \cellcolor[HTML]{D9D9D9}22.2  & 33.3                         & -                          & \cellcolor[HTML]{D9D9D9}88.9  & \cellcolor[HTML]{D9D9D9}44.4  & -                             & -                             & -                            & -                            & -                            & -                             & -                            & -                            & -                            & -                             \\
                                                    & CheeseBank\_1              & \cellcolor[HTML]{D9D9D9}0    & 0                            & \xmark                       & 0                          & \xmark                        & 0                          & 0                          & 0.5                          & \cellcolor[HTML]{D9D9D9}0.5   & 4.3                          & 0                          & \cellcolor[HTML]{D9D9D9}7.1   & \cellcolor[HTML]{D9D9D9}1.6   & 0                             & \cellcolor[HTML]{D9D9D9}33.2  & 2.2                          & \cellcolor[HTML]{D9D9D9}1.1  & 2.2                          & \cellcolor[HTML]{D9D9D9}1.6   & -                            & -                            & \cellcolor[HTML]{D9D9D9}3.3  & 0.5                           \\
                                                    & CheeseBank\_2              & \cellcolor[HTML]{D9D9D9}0    & 0                            & \xmark                       & 0                          & \xmark                        & 0                          & 0                          & 0.6                          & \cellcolor[HTML]{D9D9D9}1.1   & \cellcolor[HTML]{D9D9D9}1.1  & 0                          & \cellcolor[HTML]{D9D9D9}10.2  & \cellcolor[HTML]{D9D9D9}2.8   & 0                             & \cellcolor[HTML]{D9D9D9}7.9   & 0                            & \cellcolor[HTML]{D9D9D9}0    & 0                            & \cellcolor[HTML]{D9D9D9}0     & -                            & -                            & \cellcolor[HTML]{D9D9D9}0    & 0                             \\
                                                    & CheeseBank\_3              & \cellcolor[HTML]{D9D9D9}0    & 0                            & \xmark                       & 0                          & \xmark                        & 0                          & 0                          & 0                            & \cellcolor[HTML]{D9D9D9}1.8   & 1.8                          & 0                          & \cellcolor[HTML]{D9D9D9}10.8  & \cellcolor[HTML]{D9D9D9}1.8   & 0                             & 7.8                           & 3                            & 0                            & 3                            & \cellcolor[HTML]{D9D9D9}0     & -                            & -                            & 3                            & 3.6                           \\
                                                    & InverseFi                  & 0                            & 0                            & 0                            & 0.1                        & 0.1                           & 0                          & 0                          & 0.3                          & 0.1                           & 0                            & 0                          & 0                             & 0                             & 1.9                           & \cellcolor[HTML]{D9D9D9}16.3  & 1.5                          & \cellcolor[HTML]{D9D9D9}0    & 1.3                          & \cellcolor[HTML]{D9D9D9}0.1   & -                            & -                            & \cellcolor[HTML]{D9D9D9}2    & 0.2                           \\
                                                    & CreamFi1\_1                & 0                            & 0                            & $\emptyset$                  & 0                          & $\emptyset$                   & 0                          & 0                          & $\emptyset$                  & $\emptyset$                   & $\emptyset$                  & 0                          & -                             & -                             & -                             & -                             & -                            & -                            & -                            & -                             & -                            & -                            & $\emptyset$                  & $\emptyset$                   \\
                                                    & CreamFi1\_2                & \cellcolor[HTML]{D9D9D9}0.5  & 0.5                          & 0.1                          & 0.5                        & 0                             & \xmark                     & \xmark                     & 0.5                          & 0.1                           & 0.3                          & 0                          & -                             & -                             & 0                             & 0                             & 0                            & 0                            & 0.1                          & 0.2                           & -                            & 0                            & 0                            & 0                             \\
                                                    & CreamFi2\_1                & \cellcolor[HTML]{D9D9D9}0    & 0.3                          & 0                            & 0.3                        & 0.3                           & 0                          & 0                          & 0.8                          & \cellcolor[HTML]{D9D9D9}1.1   & \cellcolor[HTML]{D9D9D9}1.1  & 0                          & 0                             & 0                             & 2.9                           & \cellcolor[HTML]{D9D9D9}9.5   & 1.8                          & \cellcolor[HTML]{D9D9D9}0    & 0                            & \cellcolor[HTML]{D9D9D9}0.8   & 2.6                          & -                            & \cellcolor[HTML]{D9D9D9}0.8  & 0.5                           \\
                                                    & CreamFi2\_2                & \cellcolor[HTML]{D9D9D9}0    & 0                            & 1.5                          & 0                          & 1.9                           & 0                          & 0                          & 3.7                          & \cellcolor[HTML]{D9D9D9}3.3   & \cellcolor[HTML]{D9D9D9}2.6  & 0                          & 0                             & 0                             & 0                             & \cellcolor[HTML]{D9D9D9}0     & 0                            & \cellcolor[HTML]{D9D9D9}0    & 0                            & \cellcolor[HTML]{D9D9D9}0.4   & 2.6                          & -                            & \cellcolor[HTML]{D9D9D9}0    & 1.1                           \\
                                                    & CreamFi2\_3                & 3.8                          & 0                            & 9                            & 0                          & \cellcolor[HTML]{D9D9D9}12.8  & 0                          & 0                          & \cellcolor[HTML]{D9D9D9}7.7  & \cellcolor[HTML]{D9D9D9}3.8   & \cellcolor[HTML]{D9D9D9}2.6  & 0                          & -                             & -                             & 10.3                          & \cellcolor[HTML]{D9D9D9}0     & 3.8                          & 0                            & 3.8                          & 0                             & 0                            & -                            & 3.8                          & 1.3                           \\
                                                    & CreamFi2\_4                & 0                            & 0                            & \xmark                       & 0                          & \xmark                        & 0                          & 0                          & 0                            & \cellcolor[HTML]{D9D9D9}0     & \cellcolor[HTML]{D9D9D9}0    & 0                          & 0                             & 0                             & -                             & \cellcolor[HTML]{D9D9D9}20.7  & 0                            & \cellcolor[HTML]{D9D9D9}3.4  & 0                            & \cellcolor[HTML]{D9D9D9}3.4   & 3.4                          & -                            & \cellcolor[HTML]{D9D9D9}3.4  & 0                             \\
                                                    & RariCapital1               & 0                            & 0                            & 6                            & 0                          & 4.5                           & \cellcolor[HTML]{D9D9D9}0  & \cellcolor[HTML]{D9D9D9}0  & \cellcolor[HTML]{D9D9D9}6.5  & \cellcolor[HTML]{D9D9D9}1.5   & \cellcolor[HTML]{D9D9D9}0    & 0                          & -                             & -                             & -                             & -                             & \cellcolor[HTML]{D9D9D9}0.5  & \cellcolor[HTML]{D9D9D9}0    & -                            & -                             & -                            & \cellcolor[HTML]{D9D9D9}0.5  & \cellcolor[HTML]{D9D9D9}0.5  & 0                             \\
                                                    & RariCapital2\_1            & \cellcolor[HTML]{D9D9D9}0    & 0                            & 0                            & 0                          & 0                             & \cellcolor[HTML]{D9D9D9}0  & \cellcolor[HTML]{D9D9D9}0  & \cellcolor[HTML]{D9D9D9}1.1  & \cellcolor[HTML]{D9D9D9}1.1   & \cellcolor[HTML]{D9D9D9}0.5  & 0                          & -                             & -                             & \cellcolor[HTML]{D9D9D9}10.9  & \cellcolor[HTML]{D9D9D9}6     & \cellcolor[HTML]{D9D9D9}0.5  & \cellcolor[HTML]{D9D9D9}0.5  & 0                            & \cellcolor[HTML]{D9D9D9}0     & -                            & \cellcolor[HTML]{D9D9D9}0.5  & \cellcolor[HTML]{D9D9D9}0.5  & 0                             \\
                                                    & RariCapital2\_2            & \cellcolor[HTML]{D9D9D9}1.8  & 0                            & \cellcolor[HTML]{D9D9D9}0.4  & 0                          & 0.4                           & \cellcolor[HTML]{D9D9D9}0  & \cellcolor[HTML]{D9D9D9}0  & \cellcolor[HTML]{D9D9D9}1.8  & \cellcolor[HTML]{D9D9D9}0.4   & \cellcolor[HTML]{D9D9D9}0.4  & 0                          & -                             & -                             & \cellcolor[HTML]{D9D9D9}6.7   & \cellcolor[HTML]{D9D9D9}2.2   & \cellcolor[HTML]{D9D9D9}0.9  & \cellcolor[HTML]{D9D9D9}0    & \cellcolor[HTML]{D9D9D9}0.9  & \cellcolor[HTML]{D9D9D9}0.4   & 1.3                          & -                            & \cellcolor[HTML]{D9D9D9}0.9  & 2.7                           \\
                                                    & RariCapital2\_3            & \cellcolor[HTML]{D9D9D9}0    & 0                            & 0                            & 0                          & 0                             & 0                          & 0                          & 0                            & \cellcolor[HTML]{D9D9D9}0.8   & \cellcolor[HTML]{D9D9D9}1.7  & 0                          & -                             & -                             & 0                             & 0                             & 0                            & 0                            & 0.8                          & \cellcolor[HTML]{D9D9D9}0     & 1.7                          & -                            & 0                            & 0                             \\
                                                    & RariCapital2\_4            & \cellcolor[HTML]{D9D9D9}0    & 0                            & 0                            & 0                          & 0                             & 0                          & 0                          & 0.4                          & \cellcolor[HTML]{D9D9D9}0.4   & \cellcolor[HTML]{D9D9D9}2.6  & 0                          & -                             & -                             & 0                             & \cellcolor[HTML]{D9D9D9}0     & 0                            & \cellcolor[HTML]{D9D9D9}0    & 0                            & \cellcolor[HTML]{D9D9D9}0     & 0.9                          & -                            & \cellcolor[HTML]{D9D9D9}0    & 0                             \\
\midrule \multirow{-19}{*}{\vertical{Lending}}      & XCarnival                  & 0                            & \cellcolor[HTML]{D9D9D9}6.9  & \xmark                       & 5.9                        & \xmark                        & 0                          & 0                          & \cellcolor[HTML]{D9D9D9}2    & \cellcolor[HTML]{D9D9D9}11.8  & \cellcolor[HTML]{D9D9D9}8.8  & 0                          & -                             & -                             & 0                             & \cellcolor[HTML]{D9D9D9}4.9   & 3.9                          & 1                            & 4.9                          & \cellcolor[HTML]{D9D9D9}28.4  & 3.9                          & 3.9                          & 1                            & 1                             \\
                                                    & Harvest1                   & \cellcolor[HTML]{D9D9D9}0    & 0.5                          & \cellcolor[HTML]{D9D9D9}0    & 0                          & \cellcolor[HTML]{D9D9D9}0     & \cellcolor[HTML]{D9D9D9}0  & \cellcolor[HTML]{D9D9D9}0  & \cellcolor[HTML]{D9D9D9}0.2  & \cellcolor[HTML]{D9D9D9}0     & \cellcolor[HTML]{D9D9D9}0    & 0                          & \cellcolor[HTML]{D9D9D9}29.3  & \cellcolor[HTML]{D9D9D9}0     & \cellcolor[HTML]{D9D9D9}1.1   & -                             & 0                            & \cellcolor[HTML]{D9D9D9}0.2  & 0                            & \cellcolor[HTML]{D9D9D9}0.2   & -                            & -                            & \cellcolor[HTML]{D9D9D9}2.6  & 0.2                           \\
                                                    & Harvest2                   & \cellcolor[HTML]{D9D9D9}0    & 0                            & \cellcolor[HTML]{D9D9D9}0    & 0                          & \cellcolor[HTML]{D9D9D9}0     & \cellcolor[HTML]{D9D9D9}0  & \cellcolor[HTML]{D9D9D9}0  & \cellcolor[HTML]{D9D9D9}0    & \cellcolor[HTML]{D9D9D9}0     & \cellcolor[HTML]{D9D9D9}0    & 0                          & \cellcolor[HTML]{D9D9D9}34.7  & \cellcolor[HTML]{D9D9D9}0     & \cellcolor[HTML]{D9D9D9}0     & -                             & 0                            & \cellcolor[HTML]{D9D9D9}0    & 0.2                          & \cellcolor[HTML]{D9D9D9}0.2   & -                            & -                            & \cellcolor[HTML]{D9D9D9}1.2  & 0.2                           \\
                                                    & ValueDeFi                  & 0                            & \cellcolor[HTML]{D9D9D9}0    & \xmark                       & 0                          & \xmark                        & 0                          & \cellcolor[HTML]{D9D9D9}0  & \cellcolor[HTML]{D9D9D9}3.4  & 1.1                           & 2.3                          & 0                          & \cellcolor[HTML]{D9D9D9}64.8  & 0                             & \cellcolor[HTML]{D9D9D9}73.9  & -                             & \cellcolor[HTML]{D9D9D9}0    & \cellcolor[HTML]{D9D9D9}0    & 0                            & 0                             & -                            & -                            & \cellcolor[HTML]{D9D9D9}8    & 1.1                           \\
                                                    & Yearn1                     & 0                            & 0                            & \xmark                       & 0                          & \cellcolor[HTML]{D9D9D9}0     & \cellcolor[HTML]{D9D9D9}0  & \cellcolor[HTML]{D9D9D9}0  & \cellcolor[HTML]{D9D9D9}0    & \cellcolor[HTML]{D9D9D9}0     & 0                            & 0                          & -                             & -                             & -                             & -                             & -                            & -                            & -                            & -                             & -                            & -                            & $\emptyset$                  & $\emptyset$                   \\
                                                    & Yearn1\_I                  & \cellcolor[HTML]{D9D9D9}0    & 0                            & \xmark                       & 0                          & \xmark                        & \xmark                     & \xmark                     & \cellcolor[HTML]{D9D9D9}0    & \cellcolor[HTML]{D9D9D9}0.1   & 0.1                          & 0                          & -                             & -                             & -                             & -                             & \cellcolor[HTML]{D9D9D9}0    & \cellcolor[HTML]{D9D9D9}0.1  & \cellcolor[HTML]{D9D9D9}0.1  & 1.8                           & 0                            & -                            & \cellcolor[HTML]{D9D9D9}4.4  & 0                             \\
                                                    & VisorFi                    & \cellcolor[HTML]{D9D9D9}0    & 0                            & \xmark                       & 0                          & \xmark                        & 0                          & 0                          & 0                            & \cellcolor[HTML]{D9D9D9}0     & \cellcolor[HTML]{D9D9D9}0    & \cellcolor[HTML]{D9D9D9}0  & -                             & -                             & -                             & -                             & 0                            & \cellcolor[HTML]{D9D9D9}1.2  & 0                            & \cellcolor[HTML]{D9D9D9}0     & -                            & -                            & \cellcolor[HTML]{D9D9D9}4.2  & 0                             \\
                                                    & UmbrellaNetwork            & 0                            & 11.8                         & \xmark                       & 11.8                       & \xmark                        & 0                          & 0                          & 5.9                          & 0                             & 0                            & 0                          & -                             & -                             & -                             & -                             & 5.9                          & 0                            & 5.9                          & 0                             & -                            & -                            & 5.9                          & 17.6                          \\
\midrule \multirow{-10}{*}{\vertical{Yield-Earning}}& PickleFi                   & \cellcolor[HTML]{D9D9D9}0    & 0                            & 0                            & 0                          & 0                             & 0                          & 0                          & \cellcolor[HTML]{D9D9D9}0    & \cellcolor[HTML]{D9D9D9}0.1   & \cellcolor[HTML]{D9D9D9}0.1  & 0                          & -                             & -                             & -                             & -                             & 0                            & \cellcolor[HTML]{D9D9D9}0.1  & -                            & -                             & -                            & -                            & \cellcolor[HTML]{D9D9D9}0.1  & 0.1                           \\
                                                    & Eminence                   & 0                            & \xmark                       & \xmark                       & 0                          & \xmark                        & \cellcolor[HTML]{D9D9D9}0  & \cellcolor[HTML]{D9D9D9}0  & 0                            & \cellcolor[HTML]{D9D9D9}0.1   & \cellcolor[HTML]{D9D9D9}0    & 0                          & -                             & -                             & \cellcolor[HTML]{D9D9D9}20.5  & -                             & \cellcolor[HTML]{D9D9D9}0    & \cellcolor[HTML]{D9D9D9}0.5  & \cellcolor[HTML]{D9D9D9}0    & \cellcolor[HTML]{D9D9D9}0.1   & -                            & -                            & \cellcolor[HTML]{D9D9D9}0.4  & 0                             \\
                                                    & Opyn                       & \cellcolor[HTML]{D9D9D9}0    & 0                            & \cellcolor[HTML]{D9D9D9}0    & 0                          & \cellcolor[HTML]{D9D9D9}0     & \cellcolor[HTML]{D9D9D9}0  & \cellcolor[HTML]{D9D9D9}0  & 0                            & \cellcolor[HTML]{D9D9D9}0     & \cellcolor[HTML]{D9D9D9}0    & 0                          & -                             & -                             & -                             & -                             & -                            & -                            & -                            & -                             & -                            & -                            & $\emptyset$                  & $\emptyset$                   \\
                                                    & IndexFi                    & 0                            & 0                            & 0                            & \cellcolor[HTML]{D9D9D9}0  & 0                             & \cellcolor[HTML]{D9D9D9}0  & \cellcolor[HTML]{D9D9D9}0  & 0                            & \cellcolor[HTML]{D9D9D9}0.7   & \cellcolor[HTML]{D9D9D9}0.1  & 0                          & -                             & -                             & \cellcolor[HTML]{D9D9D9}6.6   & -                             & \cellcolor[HTML]{D9D9D9}0    & \cellcolor[HTML]{D9D9D9}0    & \cellcolor[HTML]{D9D9D9}0.1  & \cellcolor[HTML]{D9D9D9}0     & \cellcolor[HTML]{D9D9D9}0.4  & -                            & \cellcolor[HTML]{D9D9D9}0    & 0                             \\
                                                    & RevestFi                   & 0                            & 0                            & \xmark                       & 0                          & \cellcolor[HTML]{D9D9D9}3.7   & \cellcolor[HTML]{D9D9D9}0  & \cellcolor[HTML]{D9D9D9}0  & 4.1                          & \cellcolor[HTML]{D9D9D9}0.4   & \cellcolor[HTML]{D9D9D9}1.8  & 0                          & -                             & -                             & -                             & -                             & -                            & -                            & -                            & -                             & -                            & -                            & $\emptyset$                  & $\emptyset$                   \\
                                                    & RevestFi\_I                & \cellcolor[HTML]{D9D9D9}1.4  & 0                            & \cellcolor[HTML]{D9D9D9}0    & 0                          & \cellcolor[HTML]{D9D9D9}4.1   & \cellcolor[HTML]{D9D9D9}0  & \cellcolor[HTML]{D9D9D9}0  & 4.6                          & \cellcolor[HTML]{D9D9D9}0.5   & \cellcolor[HTML]{D9D9D9}2.1  & \cellcolor[HTML]{D9D9D9}0  & -                             & -                             & -                             & -                             & -                            & -                            & -                            & -                             & -                            & -                            & -                            & -                             \\
                                                    & DODO                       & 0                            & \cellcolor[HTML]{D9D9D9}0    & 8.3                          & \cellcolor[HTML]{D9D9D9}0  & \xmark                        & 0                          & 0                          & 0                            & 33.3                          & 25                           & 0                          & -                             & -                             & 0                             & -                             & -                            & -                            & -                            & -                             & -                            & -                            & $\emptyset$                  & $\emptyset$                   \\
                                                    & Punk\_1                    & \cellcolor[HTML]{D9D9D9}0    & \cellcolor[HTML]{D9D9D9}0    & $\emptyset$                  & \cellcolor[HTML]{D9D9D9}0  & $\emptyset$                   & -                          & -                          & 25                           & \cellcolor[HTML]{D9D9D9}0     & 0                            & 0                          & -                             & -                             & -                             & -                             & 0                            & 0                            & -                            & -                             & -                            & -                            & 0                            & 0                             \\
                                                    & Punk\_2                    & \cellcolor[HTML]{D9D9D9}0    & \cellcolor[HTML]{D9D9D9}0    & $\emptyset$                  & \cellcolor[HTML]{D9D9D9}0  & 0                             & -                          & -                          & \xmark                       & 0                             & 0                            & 0                          & -                             & -                             & -                             & -                             & 0                            & 0                            & -                            & -                             & -                            & -                            & 0                            & 0                             \\
                                                    & Punk\_3                    & \cellcolor[HTML]{D9D9D9}0    & \cellcolor[HTML]{D9D9D9}0    & $\emptyset$                  & \cellcolor[HTML]{D9D9D9}0  & $\emptyset$                   & -                          & -                          & 18.2                         & 0                             & 0                            & 0                          & -                             & -                             & -                             & -                             & 0                            & 0                            & -                            & -                             & -                            & -                            & 0                            & 0                             \\
                                                    & BeanstalkFarms             & 0                            & 0                            & \cellcolor[HTML]{D9D9D9}0.2  & 0                          & 0.4                           & \xmark                     & \cellcolor[HTML]{D9D9D9}0  & 0.2                          & \cellcolor[HTML]{D9D9D9}0.7   & \cellcolor[HTML]{D9D9D9}0.6  & 0                          & -                             & -                             & \cellcolor[HTML]{D9D9D9}31.8  & -                             & \cellcolor[HTML]{D9D9D9}0    & \cellcolor[HTML]{D9D9D9}0.1  & \cellcolor[HTML]{D9D9D9}0    & \cellcolor[HTML]{D9D9D9}0     & -                            & -                            & \cellcolor[HTML]{D9D9D9}0.8  & 0.3                           \\  
\multirow{-11}{*}{\vertical{Others}}                & BeanstalkFarms\_I          & \cellcolor[HTML]{D9D9D9}0    & 2.2                          & \cellcolor[HTML]{D9D9D9}0    & 2.2                        & \cellcolor[HTML]{D9D9D9}0     & -                          & -                          & 0                            & \cellcolor[HTML]{D9D9D9}0     & \cellcolor[HTML]{D9D9D9}0    & -                          & -                             & -                             & -                             & -                             & -                            & -                            & -                            & -                             & -                            & -                            & -                            & -                             \\  \bottomrule

                                                    & \# Contracts Applied       & 42                           & 39                           & 22                           & 39                         & 25                            & 33                         & 33                         & 37                           & 41                            & 41                           & 40                         & 11                            & 11                            & 21                            & 16                            & 34                           & 34                           & 28                           & 28                            & 11                           & 7                            & 33                           & 33                            \\
                                                    & \# Contracts Protected     & 23                           & 6                            & 7                            & 5                          & 9                             & 11                         & 13                         & 12                           & 30                            & 23                           & 2                          & 7                             & 6                             & 8                             & 12                            & 9                            & 22                           & 5                            & 22                            & 1                            & 2                            & 23                           & 1                             \\
                                                    & \# Hacks Blocked           & 15                           & 4                            & 6                            & 3                          & 8                             & 9                          & 11                         & 10                           & 18                            & 15                           & 2                          & 5                             & 4                             & 7                             & 6                             & 8                            & 17                           & 5                            & 15                            & 1                            & 2                            & 18                           & 1                             \\
                                                    & Average FP(\%)             & 0.2                          & 0.6                          & 1.8                          & 0.5                        & 2.6                           & 0                          & 0                          & 2.4                          & 2.4                           & 2.6                          & 0                          & 22.3                          & 4.6                           & 7.9                           & 12.6                          & 0.7                          & 0.4                          & 0.8                          & 1.4                           & 1.5                          & 0.7                          & 1.6                          & 0.9                          \\ \bottomrule

                                                \end{tabular}

\end{adjustbox}
\vspace{-3mm}
\end{table}

\newpage

\begin{table}[!htbp]
    \scriptsize
	\vspace{-2mm}
    \caption{Summarized Results of Invariants Effectiveness Evaluation.} 
    \vspace{-4mm}
	\label{tab:rq1summary}
    \tabcolsep=0.11cm
    \resizebox{\columnwidth}{!}{
    	\begin{tabular}{|l|lllll|lll|ll|l|ll|ll|llll|llll|}
			\hline
    																& \multicolumn{5}{c|}{AccessControl}                    						& \multicolumn{3}{c|}{TimeLock}  					& \multicolumn{2}{c|}{GasCtrl} 		& \multicolumn{1}{c|}{Re} 	& \multicolumn{2}{c|}{Oracle} 				& \multicolumn{2}{c|}{Storage} 		& \multicolumn{4}{c|}{Money Flow}                   				& \multicolumn{4}{c|}{Data Flow}                 					\\ \hline
    																& \cellcolor[HTML]{D9D9D9}\rot{EOA}  		& \rot{SO} 			& \rot{SM} 			& \rot{OO} 		& \rot{OM} 			& \rot{SB} 			& \cellcolor[HTML]{D9D9D9}\rot{OB} 		& \rot{LU} 			& \cellcolor[HTML]{D9D9D9}\rot{GS}    	& \rot{GC}       	& \rot{RE}         			& \rot{OR}     	& \rot{OD}    				& \rot{TSU}     & \rot{TBU}    		& \rot{TIU} 	& \cellcolor[HTML]{D9D9D9}\rot{TOU} 	& \rot{TIRU} 	& \rot{TORU} 	& \rot{MU} 		& \rot{CVU} 	& \cellcolor[HTML]{D9D9D9}\rot{DFU} 	& \rot{DFL} 						\\ \hline
			\# Contracts Applied(42)   								& \cellcolor[HTML]{D9D9D9}42 		 		& 39  				& 22  				& 39  			& 25  				& 33 				& \cellcolor[HTML]{D9D9D9}33 			& 37  				& \cellcolor[HTML]{D9D9D9}41  			& 41  				& 40 						& 11   			& 11  						& 21  			& 16   				& 34  			& \cellcolor[HTML]{D9D9D9}34  			& 28  			& 28  			& 11  			& 7   			& \cellcolor[HTML]{D9D9D9}33  			& 33  											\\
			\# Contracts Protected(42) 								& \cellcolor[HTML]{D9D9D9}23  				& 6   				& 7   				& 5   			& 9  				& 11 				& \cellcolor[HTML]{D9D9D9}13 			& 12  				& \cellcolor[HTML]{D9D9D9}30  			& 23  				& 2  						& 7    			& 6   						& 8   			& 12   				& 9   			& \cellcolor[HTML]{D9D9D9}22  			& 5   			& 22  			& 1   			& 2   			& \cellcolor[HTML]{D9D9D9}23  			& 1   											\\
			\# \textbf{Hacks Blocked(27)}     						& \cellcolor[HTML]{D9D9D9}\textbf{15}  		& \textbf{4}   		& \textbf{6}   		& \textbf{3}   	& \textbf{8}   		& \textbf{9}  		& \cellcolor[HTML]{D9D9D9}\textbf{11} 	& \textbf{10}  		& \cellcolor[HTML]{D9D9D9}\textbf{18}  	& \textbf{15}  		& \textbf{2}  				& \textbf{5}    & \textbf{4}   				& \textbf{7}    & \textbf{6}    	& \textbf{8}   	& \cellcolor[HTML]{D9D9D9}\textbf{17}  	& \textbf{5}   	& \textbf{15}  	& \textbf{1}   	& \textbf{2}   	& \cellcolor[HTML]{D9D9D9}\textbf{18}  	& \textbf{1}   									\\
			Average FP(\%)             								& \cellcolor[HTML]{D9D9D9}0.2 				& 0.6 				& 1.8  				& 0.5 			& 2.6 				& 0  				& \cellcolor[HTML]{D9D9D9}0  			& 2.4 				& \cellcolor[HTML]{D9D9D9}2.4 			& 2.6 				& 0  						& 22.3 			& 4.6 						& 7.9 			& 12.6				& 0.7 			& \cellcolor[HTML]{D9D9D9}0.4 			& 0.8 			& 1.4  			& 1.5 			& 0.7 			& \cellcolor[HTML]{D9D9D9}1.6 			& 0.9 											\\ \hline
    	\end{tabular}
    }
	\vspace{-4mm}
\end{table}



\textbf{Results.} \Cref{tab:rq1} provides a detailed breakdown of the effectiveness evaluation for 
each invariant applied across the $42$ contracts. 
The table lists the number of contracts on which an invariant is applied, 
the number of contracts successfully protected by the invariants, 
the number of hacks blocked, and the false positive rate for each invariant. 
The row \textbf{\# Hacks Blocked} stands out as the most crucial metric as it 
directly measures the capability of each invariant to block exploits.
The average false positive rate is also an important metric as it quantifies 
the potential impact on regular users, reflecting the trade-off between security and usability.

\Cref{tab:rq1summary} provides a comprehensive summary of the effectiveness
evaluation for the invariants applied across various contracts. The table lists several
key metrics: the number of contracts on which an invariant is applied, the number of
contracts successfully protected by the invariants, the number of hacks blocked, and the average
false positive (FP) rate. Among these, the row \textbf{\# Hacks Blocked} stands out as the most
crucial metric as it directly measures the capability of each invariant to block exploits. 
The average FP rate is also an important metric as it quantifies the potential impact on
regular users, reflecting the trade-off between security and usability.

The applicability of the invariants varies across different categories. Access Control, Time Lock,
Gas Control, Money Flow, and Data Flow are universally applicable, protecting a broad range of
contracts and blocking numerous hacks. On the contrary, categories such as ReEntrancy, Oracle,
and Storage have narrower scopes, applicable only to specific types of contracts. For
instance, the ReEntrancy invariant we studied is effective only against common single-contract
reentrancy attacks. Other attack types, such as read-only or cross-contract reentrancy seen in
CreamFi1, RariCapital1, and RariCapital2, require more specialized invariants and are left as
future work.

For true positives, in each category of Access Control, Time Lock, Gas Control, Data Flow, and Money Flow, there is a standout invariant that proves most effective at blocking hacks: EOA for Access Control, OB for Time Lock, GS for Gas Control, TOU for Money Flow, and DFU for Data Flow. These invariants block the highest number of hacks in their respective categories.

For false positives, EOA, OB and TOU have an average FP rate below 0.4\%, while GS and DFU have a low FP rate below 2.4\%.  This low rate indicates that these
invariants have a small impact on regular user transactions, thereby making them practical for real-world deployment. The elevated false positive rates observed for OR, TSU, and TBU are primarily because of the fluctuating nature of oracle values, total supply, and total borrow. Using upper-bound or range-based invariants for these categories could inadvertently block all transactions once these values exceed a certain threshold.

\vspace{-1mm}
\begin{tcolorbox}[width=\linewidth, 
	left=0mm, right=0mm, top=0mm, bottom=0mm, 
	opacityfill=0.1]
	\textbf{Answer to RQ1}:
	EOA in Access Control, OB in Time Lock, GS in Gas Control, TOU in Money Flow, and DFU in Data Flow are the most effective in blocking hacks with low false positive rates.
\end{tcolorbox}
\vspace{-3mm}

\subsection{RQ2: Study of False Positives and True Positives}
\label{sec:rq2}

\textbf{Case Studies.} In our second research question (RQ2), we explore the
bypassability of the invariants for both malicious hackers and normal users. Hackers will be informed when the invariants are deployed in the target contract from the source code or the bytecode. It is natural to ask whether malicious hackers can bypass the invariants and still gain profit if they realize the existence of such invariant guards. We manually analyze every exploit transaction blocked by each invariant (true positives) to check its bypassability. For each case, we assign one of three categories: \textbf{C1:} the exploit is entirely blocked, and the hacker can no longer gain any profit; \textbf{C2:} the exploit is partially blocked, resulting in significantly reduced profits for the hacker; and \textbf{C3:} the hacker can still achieve profits similar to historical data, with some adjustments to their exploit code.

We also manually analyze the false positives generated by our invariants to assess their impact on
regular users. For this, we sample up to 10 transactions from the false positives for each
invariant and evaluate their bypassability. We operate under the assumption that regular
users have the option to split transactions with large parameters into transactions with 
smaller parameters, lower
the gas for their transactions, or simply wait for some time to transact 
again after their transaction is blocked. 
Based on these criteria, we categorize the false positives into three groups: \textbf{D1:} 
transaction is completely blocked and cannot be bypassed through simple means; \textbf{D2:} 
transaction can be bypassed by breaking it down into smaller transactions; and \textbf{D3:} users
can bypass the transaction by reducing the gas or waiting for some time.
For both true and false positives,
two authors independently labeled the bypassability results with the third author to resolve divergence of views.

\begin{table}[!htbp]
    \scriptsize
    \vspace{-2mm}
    \caption{Bypassability Results of Hacks Blocked (TPs) and Sampled Normal Transactions Blocked (FPs).}
    \label{tab:bypass}
    \vspace{-4mm}
    \scriptsize
    \tabcolsep=0.11cm

    \begin{tabular}{|l|l|lllll|lll|ll|l|ll|ll|llll|llll|}
        \hline
                                                &    & \multicolumn{5}{c|}{Access Control}                    & \multicolumn{3}{c|}{Time Lock}  & \multicolumn{2}{c|}{GasCtrl} & \multicolumn{1}{c|}{Re} & \multicolumn{2}{c|}{Oracle} & \multicolumn{2}{c|}{Storage}                                                                                                                                                                                                                                                                      & \multicolumn{4}{c|}{Money Flow}                      & \multicolumn{4}{c|}{Data Flow}                         \\ \hline
                                                &    & \cellcolor[HTML]{D9D9D9}\rot{EOA}                     & \rot{SO}                     & \rot{SM}                     & \rot{OO}                     & \rot{OM}                     & \rot{SB}                     & \cellcolor[HTML]{D9D9D9}\rot{OB}                     & \rot{LU}                     & \cellcolor[HTML]{D9D9D9}\rot{GS}                     & \cellcolor[HTML]{D9D9D9}\rot{GC}                     & \rot{RE}                     & \rot{OR}                     & \rot{OD}                     & \rot{TSU}                     & \rot{TBU}                       & \rot{TIU} & \cellcolor[HTML]{D9D9D9}\rot{TOU} & \rot{TIRU} & \rot{TORU}    & \rot{MU} & \rot{CVU} & \cellcolor[HTML]{D9D9D9}\rot{DFU} & \rot{DFL}          \\   \hline
    \multirow{3}{*}{\vertical{TPs}}             & C1 & \cellcolor[HTML]{D9D9D9}15                            & 4                            & 6                            & 3                            & 8                            & 2                            & \cellcolor[HTML]{D9D9D9}10                           & 5                            & \cellcolor[HTML]{D9D9D9}0                            & \cellcolor[HTML]{D9D9D9}15                           & 2                            & 3                            & 2                            & 2                             & 3                               & 7         & \cellcolor[HTML]{D9D9D9}12        & 2          & 7             & 1        & 1         & \cellcolor[HTML]{D9D9D9}12        & 1                  \\
                                                & C2 & \cellcolor[HTML]{D9D9D9}0                             & 0                            & 0                            & 0                            & 0                            & 0                            & \cellcolor[HTML]{D9D9D9}0                            & 0                            & \cellcolor[HTML]{D9D9D9}1                            & \cellcolor[HTML]{D9D9D9}0                            & 0                            & 2                            & 2                            & 5                             & 3                               & 0         & \cellcolor[HTML]{D9D9D9}0         & 2          & 8             & 0        & 0         & \cellcolor[HTML]{D9D9D9}0         & 0                  \\
                                                & C3 & \cellcolor[HTML]{D9D9D9}0                             & 0                            & 0                            & 0                            & 0                            & 7                            & \cellcolor[HTML]{D9D9D9}1                            & 5                            & \cellcolor[HTML]{D9D9D9}17                           & \cellcolor[HTML]{D9D9D9}0                            & 0                            & 0                            & 0                            & 0                             & 0                               & 1         & \cellcolor[HTML]{D9D9D9}5         & 1          & 0             & 1        & 1         & \cellcolor[HTML]{D9D9D9}6         & 0                  \\  \hline
    \multirow{3}{*}{\vertical{FPs}}             & D1 & \cellcolor[HTML]{D9D9D9}10                            & 10                           & 10                           & 10                           & 10                           & 0                            & \cellcolor[HTML]{D9D9D9}0                            & 1                            & \cellcolor[HTML]{D9D9D9}0                            & \cellcolor[HTML]{D9D9D9}10                           & 0                            & 10                           & 10                           & 10                            & 10                              & 0         & \cellcolor[HTML]{D9D9D9}0         & 0          & 1             & 10       & 0         & \cellcolor[HTML]{D9D9D9}4         & 10                 \\
                                                & D2 & \cellcolor[HTML]{D9D9D9}0                             & 0                            & 0                            & 0                            & 0                            & 0                            & \cellcolor[HTML]{D9D9D9}0                            & 0                            & \cellcolor[HTML]{D9D9D9}0                            & \cellcolor[HTML]{D9D9D9}0                            & 0                            & 0                            & 0                            & 0                             & 0                               & 10        & \cellcolor[HTML]{D9D9D9}10        & 10         & 9             & 0        & 8         & \cellcolor[HTML]{D9D9D9}6         & 0                  \\
                                                & D3 & \cellcolor[HTML]{D9D9D9}0                             & 0                            & 0                            & 0                            & 0                            & 0                            & \cellcolor[HTML]{D9D9D9}0                            & 9                            & \cellcolor[HTML]{D9D9D9}10                           & \cellcolor[HTML]{D9D9D9}0                            & 0                            & 0                            & 0                            & 0                             & 0                               & 0         & \cellcolor[HTML]{D9D9D9}0         & 0          & 0             & 0        & 0         & \cellcolor[HTML]{D9D9D9}0         & 0                  \\  \hline
    \end{tabular}
    \vspace{-3mm}
    \end{table}

\textbf{Results.} \Cref{tab:bypass} offers a comprehensive view of how both attackers 
and normal users can potentially bypass the invariant guards. The table is 
divided into two major rows: True Positives (TPs) and False Positives (FPs). 
For TPs, we have further categorized the effectiveness of the invariant guards as C1, C2, and C3, 
signifying the level of success the hacker has in bypassing the guard. 
For FPs, we use tags D1, D2, and D3 to illustrate how easily normal users can circumvent these guards.

Some invariant categories can easily be bypassed by both hackers and normal users. For example, GS,
though it blocks the most hacks in RQ1, can be easily bypassed by changing the gas passed to the
specific function call of the target contract.
Some other invariants, such as GC, cannot be bypassed by either hackers or normal users.
EOA also fall into this category. OB interestingly has no false
positives, but in practice it is very hard to be bypassed by normal users.

For certain invariants such as TOU and DFU, there is a dichotomy where they block exploit
transactions effectively, but normal users can still bypass them.
They make certain exploits impossible by preventing hackers from reaching specific contract states
required for profitability.
Regular users, who usually do not attempt to manipulate contract states for exploitative gains, can
often bypass these invariants by simply splitting their larger transactions into smaller ones.
Though this results in more transactions and higher gas costs, it does not impede their primary objectives.



\vspace{-1mm}
\begin{tcolorbox}[width=\linewidth, 
	left=0mm, right=0mm, top=0mm, bottom=0mm, 
	opacityfill=0.1]
	\textbf{Answer to RQ2}:
	Most of invariants behave similarly for both hackers and normal users, either being easily
	bypassed (such as GS) or not bypassable at all (such as EOA, OB, GC). Some invariants
	could block hackers while allowing normal users to circumvent them (such as TOU and DFU).
\end{tcolorbox}
\vspace{-3mm}

\subsection{RQ3: Effectiveness of Combination of Invariants}
\label{sec:rq3}

\textbf{Experiment.}
\label{subsec:eva:rq3}
In RQ3, we look into the effectiveness of various invariant combinations in safeguarding smart
contracts.
Building on insights from RQ1 and RQ2, we identify a set of five effective
invariants (EOA, SO, GC, TOU, and DFU), replacing GS with GC considering that it can be easily
bypassed.
Our aim is to explore whether these invariants, each effective in its own domain, can be combined
to provide a more robust defense against hacks while maintaining a low FP
rate.

After investigating the exploits blocked by each of these five invariants in RQ1, we have observed that (1) EOA
and GC can uniquely block 2 and 1 hacks, respectively; (2) no exploits can uniquely be blocked by
OB, TOU or DFU;
(3) all exploits blocked by MFU can be blocked by DFU.

We then designed an experiment that combines four invariants---EOA, GC, OB, and DFU---each chosen
for its efficacy in blocking hacks and resistance to bypass.
We consider two logical operators: conjunction ($\land$) and disjunction ($\lor$).
By enumerating all logical combinations of invariants up to a length of $4$, we can evaluate their
collective True Positives (TPs) and False Positives (FPs) based on two metrics: (1) their ability to block the maximum number of hacks (2) their ability to block the maximum number of hacks while maintaining a false positive rate below 1\%.



\textbf{Results.}
\vspace{-3mm}
\begin{table}[!htbp]
\caption{Validation Results of the Best Combined Invariants.}\label{tab:rq3}
\vspace{-3mm}
\footnotesize
    \centering
    \begin{tabular}{|ll|ll|ll|ll|}
    	\hline

    \multicolumn{4}{|c|}{EOA $\land$ GC $\land$ DFU} & \multicolumn{4}{c|}{EOA $\land$ (OB $\lor$ DFU)} \\ \hline
\# Contracts Applied     & 42    & C1  & 20  & \# Contracts Applied     & 42     & C1   & 18   \\
\# Contracts Protected   & 35    & C2  & 0   & \# Contracts Protected   & 31     & C2   & 0    \\
\# Hacks Blocked         & 23    & C3  & 3   & \# Hacks Blocked         & 20     & C3   & 2    \\
Average FP rate(\%)      & 3.99  & D1  & 8   & Average FP rate(\%)      & 0.28   & D1   & 10   \\
                         &       & D2  & 2   &                          &        & D2   & 0    \\
                         &       & D3  & 0   &                          &        & D3   & 0    \\  \hline

    \end{tabular}
    \vspace{-3mm}
    \end{table}
Table~\ref{tab:rq3} shows the validation results of the best combined invariants based on metrics (1) and (2), using the same evaluation methodology as described in RQ1 and RQ2.

The combination EOA $\land$ GC $\land$ DFU emerges as the most effective one according to the
metric (1). This result is intuitive, as this combined invariant effectively reverts a transaction
when any one of EOA, GC, or DFU is violated. Our early observation that EOA and GC can uniquely block 2
and 1 hacks, also justifies their inclusion in this composite invariant. However, this comes at the
expense of a higher false positive rate of 3.99\%.

The combination EOA $\land$ (OB $\lor$ DFU) maintains an impressively low false positive rate of
0.28\%, making it superior based on the metric (2). Its efficacy is due in part to OB's inherent
low false positive rate and to the complementary nature of OB and DFU in the context of token
transfer functions.
Both combined invariants are better at stopping hacks than individual invariants.
This is because they can work together in different parts of the smart contract code, making it
more likely they will catch harmful actions. 
Sometimes a function in the contract may only be protected by one invariant, if other invariants are not applicable to this function. In those cases, both hackers and normal users may still find a way to bypass the security measures.
\vspace{-1mm}
\begin{tcolorbox}[width=\linewidth, 
	left=0mm, right=0mm, top=0mm, bottom=0mm, 
	opacityfill=0.1]
	\textbf{Answer to RQ3}:
	The invariant guards studied and generated by \name are complimentary and their combinations are promising to be more effective on contract protection and hack prevention with lower false positive rates.
\end{tcolorbox}
\vspace{-3mm}

\subsection{RQ4: Gas Overhead of Invariant Guards}
\label{sec:rq4}

\textbf{Experiment.}
\label{subsec:eva:rq4}
In RQ4, we examine the gas overhead incurred by the deployment of individual and combined invariants. We select four benchmark contracts that represent different kinds of protocols and programming languages. The target smart contract is instrumented with the generated invariants studied in RQ3 and then compiled by either a Solidity or Vyper compiler. This process is carried out for both the original and the instrumented versions of the contract. The compiler returns an estimate of the gas consumption for each function in the contract, allowing us to calculate the gas overhead for each inserted invariant by comparing the two versions.
Subsequently, we replay all transactions within the test set on these instrumented contracts. For each transaction, we add the corresponding gas overhead when the transaction reaches where the invariants are inserted.

The total gas overhead is calculated as,
$\frac{\text{Total Gas After Instrumentation} - \text{Total Gas Before Instrumentation}}{
{\text{Total Gas Before Instrumentation}}}.$
This provides a quantitative measure of the computational burden imposed by the invariant guards, aiding in the cost-benefit analysis of their deployment.

\vspace{-3mm}
\begin{table}[!htbp]
    \scriptsize
    \caption{Runtime Gas Overhead (\%) of Different Types of Invariant Guards.}\label{tab:gas}
    \vspace{-3mm}
    \label{tab:rq4gas}

    \setlength{\tabcolsep}{4pt}
    \begin{tabular}{|l|l|l|cccccc|}
    \hline
 Kind           &  Benchmark      & Compiler        & EOA  & OB$^{*}$   & GC   & DFU  & EOA $\land$ GC $\land$ DFU & EOA $\land$ (OB $\lor$ DFU) \\   \hline
 Bridge         &  HarmonyBridge  & Solidity 0.5.17 & 0.04 & 0.59 \camera{$\rightarrow$ 0.12$^+$} & 0.02 & 0.01 & 0.07  & 0.63 \camera{$\rightarrow$ 0.16}                       \\
 Lending        &  Harvest1       & Solidity 0.5.18 & 0.00 & 0.96 \camera{$\rightarrow$ 0.02}     & 0.01 & 0.01 & 0.02  & 0.97 \camera{$\rightarrow$ 0.04}                      \\
 Yield-Earning  &  CreamFi2\_1    & Solidity 0.5.17 & 0.00 & 0.14 \camera{$\rightarrow$ 0.01}     & 0.01 & 0.00 & 0.01  & 0.15 \camera{$\rightarrow$ 0.01}                      \\
 Others         &  BeanstalkFarms & Vyper 0.2.8     & 0.00 & 4.09 \camera{$\rightarrow$ 0.68}     & /    & 0.00 & 0.00  & 4.09 \camera{$\rightarrow$ 0.68}                      \\  \hline
    \end{tabular}
    \flushleft
    \scriptsize
    \camera{$^*$: OB is relatively gas-expensive due to the use of SLOAD and SSTORE operations. We optimize this by omitting OB for transactions initiated by an EOA, which is also validated on the fly.}

    \scriptsize
    \camera{$^+$: The Ethereum Cancun upgrade on March 13, 2024, implemented EIP-1153~\cite{EIP1153}, which added TLOAD and TSTORE opcodes. These can replace SLOAD and SSTORE in OB invariant to further reduce its gas overhead. As of May 1, 2024, Solidity and Vyper compilers have not fully supported EIP-1153; estimates are based on potential opcode replacements in OB invariant. See updated gas overheads after $\rightarrow$.}
    \vspace{-4mm}
\end{table}

\textbf{Results.}
\Cref{tab:gas} lists four benchmark contracts and shows the runtime gas overhead incurred by the application of various individual and combined invariant guards. Specifically, the OB invariant introduces the highest gas overhead among all individual invariants. This is attributable to the fact that OB utilizes a new contract state variable in storage to store its hash and, therefore, necessitates a storage load or store each time it is executed. In contrast, other invariants do not require additional storage variable access. DFU has the least impact on gas overhead, largely because it merely sets an upper bound on already accessed values, whereas other invariants typically fetch opcode results for comparison. The combined invariants do not have a gas overhead significantly higher than the individual invariants. This is because the combined invariants do not access new variables, but rather utilize the existing variables to perform additional comparisons.



\vspace{-1mm}
\begin{tcolorbox}[width=\linewidth, 
	left=0mm, right=0mm, top=0mm, bottom=0mm, 
	opacityfill=0.1]
	\textbf{Answer to RQ4}:
	The gas overheads of these invariant guards are as low as 0\% - 0.68\%.
\end{tcolorbox}
\vspace{-2mm}

%
%

\subsection{\revise{RQ5: Comparative Analysis with Other State-of-the-art (SOTA) Tools}}
\label{sec:rq5}


\noindent \textbf{Experiment 1: Compare with InvCon+, a SOTA invariant mining tool.}
InvCon+~\cite{liu2024automated},
a direct follow-up work of InvCon~\cite{liu2022invcon},
leverages transaction pre/post conditions to generate invariants aimed at mitigating real-world smart contract vulnerabilities.
InvCon only infers likely invariants and only produces raw results from Daikon~\cite{ernst2007daikon}. 
In contrast, InvCon+
generates accurate invariants that are verified against the contract's transaction history.
Similar to {\name}, InvCon+ takes a target contract and its transaction history as inputs and automatically generates invariants.
We obtained InvCon+~\cite{invconplusArtifact} from its authors, and applied it on our benchmarks in Table~\ref{tab:benchmarks}.
Following the same methodology of RQ1 and RQ3, we used $70\%$ of the transaction history for training, 
testing the generated invariants on the remaining $30\%$.

\noindent \textbf{Experiment 2: Compare with TxSpector, a SOTA transaction attack detection tool.}
Though {\name} is not primarily designed for transaction anomaly detection, we explored its capability to identify attack transactions. 
TxSpector~\cite{zhang2020txspector} identifies attacks using
eight detectors: re-entrancy, unchecked call, failed send, timestamp dependency, unsecured balance, misuse of origin, suicidal, and gas-related re-entrancy.
We obtained TxSpector from its public repository~\cite{txspectorArtifact}. \camera{TxSpector operates with a modified Geth archive node to produce transaction traces that are fed into the detectors. Using components of {\name} along with an unmodified Geth archive node, we generated equivalent transaction traces for the testing sets of our benchmarks, and applied TxSpector detectors to analyze them. We also patch TxSpector to support new EVM opcodes introduced after its last update.} Consistent with the setup in the TxSpector paper~\cite{zhang2020txspector}, we set a timeout of $60$ seconds for all benign transactions. We extended this to $2$ hours for hack transactions, which are typically more complex, when applying the detectors.



\begin{table}[!htbp]
    \vspace{-3mm}
    \caption{\revise{Comparison among {\name}, InvCon+, and TxSpector. (TxSpector only takes transactions as input, thus it does
    not have statistics about contracts.)}}
    \vspace{-3mm}
    \label{tab:rq5}
    \scriptsize
    \centering
    \begin{tabular}{|l|l|l|l|}
    \hline
                                           & {\name}        & InvCon+     & TxSpector     \\ \hline
    \# Contracts Applied(42)               & 42             & 27          & -             \\
    \# Contracts Protected(42)             & 31             & 8           & -             \\
    \# Hacks(27)                           & 20 Blocked     & 3 Blocked   & 7 Detected    \\
    Average \# Invariants per Contract     & 12             & 2054        & -             \\
    Average FP rate(\%)                    & 0.28           & 73.55       & 15.30         \\ \hline
    \end{tabular}
    \vspace{-4mm}
    \end{table}

\revise{\textbf{Results.}
As shown in Table~\ref{tab:rq5}, InvCon+ was applied to $27$ contracts. The
other $15$ contracts in our benchmarks, which utilize a proxy-implementation
pattern as described in Section~\ref{sec:example}, could not be processed by
InvCon+ due to its requirement for contract logic and storage to be unified.
InvCon+ encountered errors for $16$ benchmarks when processing their
transaction histories, and successfully generated invariants for $11$ victim
contracts. These invariants secured $8$ contracts across $3$ hack incidents
(Cheesebank, Punk, Warp). However, the enhanced security comes at a substantial
cost: an average of $2054$ invariants per contract and a false positive rate of
73.55\% if directly applying all of the invariants. This makes InvCon+ not
suitable for practical application without significant human efforts to filter
out unproductive invariants.}

\revise{
TxSpector correctly identified $7$ out of $27$ transactions as malicious. 
For the remaining 20 hack transactions, TxSpector experienced a timeout on 1 
transaction and failed to flag 19 transactions. 
Although TxSpector reliably identifies single contract re-entrancy attacks, 
it struggles with detecting other attack types such as cross-contract re-entrancy and oracle manipulation. 
Additionally, it inaccurately marked $15.30$\% of benign transactions as malicious, compromising its real-world utility.}

\revise{
In contrast, {\name}, utilizing the invariant template EOA $\land$ (OB $\lor$ DFU), 
effectively secured $31$ of $42$ victim contracts across $20$ hacks
with a 
remarkably low FP rate of just $0.28$\%. 
Moreover, {\name} demonstrated enhanced practicality by generating an average of only $12$ invariants per contract, significantly outperforming InvCon+ in terms of real-world viability. 
Unlike TxSpector, which only detects hacks, {\name} not only blocks a greater number of hacks but also achieves this with a significantly lower FP rate, indicating its effectiveness in anomaly detection as well.
}




\revise{
\textbf{Identifying New Exploits:}
In the development of {\name}, we surprisingly found two previously unreported exploit transactions, earlier than any reported 
exploit transactions against
RariCapital1~\cite{RariCapital1}
and
Yearn~\cite{Yearn}. }
\revise{
The two transactions were initially reported by {\name} as false positives, because they were not flagged as hacks  
when we collected benchmarks.
However, after manual investigation, we found the two transactions caused a huge financial loss and the addresses
of the originators of the two transactions are flagged on EtherScan as ``Rari Capital Exploiter''
and ``Yearn (yDai) Exploiter'',
respectively. Thus, we believe they are indeed exploit transactions.
This discovery underscores {\name}'s potential in unveiling new exploits.}

\vspace{-1mm}
\begin{tcolorbox}[width=\linewidth,
	left=0mm, right=0mm, top=0mm, bottom=0mm,
	opacityfill=0.1]
    \revise{
	\textbf{Answer to RQ5}:
	{\name} outperforms current SOTA works on smart contract invariant mining and transaction attack detection in terms of
    both practicality and accuracy. }
\end{tcolorbox}
\vspace{-2mm}

\subsection{Threats to Validity}

The \emph{internal} threat to validity mainly lies in human mistakes in the study. Specifically,
when analyzing the possibilities of bypassing invariant guards, we may miss some possible bypassing
strategies.
To mitigate this threat, two of the authors independently labeled the results, and whenever a
conflict arises, it was resolved by the third author.
All authors have more than two years' smart contract security analysis experience.

The \emph{external} threat to validity lies in the subject selection of our study.
The type of hacks studied in our experiments may be limited and biased.
To mitigate this issue, we systematically collected all the well-known hacks from a diverse set of
sources and finally included $27$ representative hacks in our benchmark.
These attacks attribute to many different root causes, including compromised keys, hash collisions,
oracle manipulation, etc.
Their affected contracts are from diverse application domains, e.g., bridges, lending, and
yield-earning.
Therefore, we believe they are representative and can be used to evaluate the effectiveness of the
invariant guards.

\vspace{-2ex}
\section{Discussion}

In this section, we discuss some key takeaways of this work and their implications in the context of invariant-based security measures for decentralized finance (DeFi) protocols.

\noindent \textbf{Choosing Complementary Invariants.}
Our findings underscore the importance of selecting a diverse set of invariants to safeguard smart contracts. Each type of invariant serves as a unique line of defense against abnormal transactions.
For instance, invariants in time lock category act as temporal barriers, making it difficult for attackers to execute key functions like \emph{withdraw} multiple times within one transaction. On the other hand, invariants in data and money flows categories limit the token amounts that can be withdrawn in one function call. By employing a combination of these invariants, developers force attackers to only withdraw a controlled amount of tokens per transaction, which may disrupt the mechanism the attack relies on, thereby blocking the attack.

\noindent \textbf{Dynamic Parameter Updates for Invariants.}
Another key insight is the need for dynamic parameter updates for certain types of invariants. For invariants that are tied to variables that change over time—like oracle prices or storage values—parameters can quickly become obsolete. If such an invariant is violated, it could lead to a cascade of failed transactions, causing a high FP rate. Thus, it is crucial for developers to monitor these variables and adjust the invariant parameters. Conversely, for invariants related to independent actions like token transfers, the parameters can remain relatively stable, as user behavior in these domains tends to be stable over time.

\noindent \textbf{Mitigating Flash Loan Attacks.}
Our benchmarks indicate a significant prevalence of flash loan-based hacks, with $17$ of $27$ examined hacks leveraging flash loan. Flash loans enable users to borrow large amounts of tokens for the duration of one transaction, providing attackers with substantial resources to execute complex hacks. Our approach can effectively mitigate the risk posed by flash loans. Invariants like EOA and OB block flash loan hacks by enforcing attackers to split their transaction into multiple ones. Without flash loan, the attacker would need to use their own assets to execute the hack with the uncertainty of other bots' back-running between the hackers' transactions. This raise not only the technical but also financial barriers to successful attacks.

\noindent \revise{
\textbf{Impact of Invariants on Contract Composability.}
Incorporating invariant guards into smart contracts may limit their
adaptability and integration with other DeFi protocols. However, our study in
Section~\ref{sec:invariants} reveals that many invariants stem from existing
DeFi protocols requirements, underscoring the preference of developers for
security benefits over flexibility. Moreover, our findings in RQ2, as discussed
in Section~\ref{sec:evaluation}, show that certain invariants, e.g., OB,
have almost no FP, while others, e.g., TOU and DFU, can possibly block malicious transactions while allowing normal users to circumvent them. These insights imply that with careful selection and
understanding of user behaviors, developers can devise invariant guards
that minimally affect contract composability.
}


\vspace{-5mm}
\section{Related Work}

\label{sec:relatedWork}

\noindent
\textbf{Smart Contract Invariants.} 
Prior works have studied various smart contract invariants. \citeauthor{zhou2020ever}~\cite{zhou2020ever} introduces six invariants to defend against different hacks. Cider~\cite{liu2022learning} uses deep reinforcement learning to learn invariants that prevent arithmetic overflows, while SPCon~\cite{liu2022finding} recovers likely access control models from function callers in past transactions. Over~\cite{deng2024safeguarding} infers safety constraints on oracles using contract source code and oracle update history. In contrast, {\name} explores a broader range of invariants and assesses their effectiveness and bypassability against real-world attacks.

\noindent
\textbf{Smart Contract Security Analysis.}
There is a large body of works on the detection of security
vulnerabilities in smart contracts.
Oyente~\cite{luu2016making} is one of the first symbolic execution-based security tools to detect \textit{reentrancy}, \textit{mishandled exception}, \textit{transaction order dependency}, and \textit{timestamp dependency}.
It has been extended to detect \textit{greedy}, \textit{prodigal} and \textit{suicidal}
contracts~\cite{nikolic2018finding}.
Other well-known symbolic-execution tools include Manticore~\cite{manticore} and
Mythril~\cite{mythril} which are able to find other types of vulnerabilities, e.g.,
\textit{dangerous delegatecall}, \textit{integer overflow}, etc.
Slither~\cite{slither} is another popular static security analysis tool for smart contracts.
It performs data flow and control flow dependency analysis to support up to $87$ bug detectors.
Other static analyzers include SmartCheck~\cite{tikhomirov2018smartcheck} mainly targeting
\textit{bad coding practices}, Securify~\cite{securify} and Ethainter~\cite{brent2020ethainter}
for finding \textit{information-flow} vulnerabilities.
Moreover, dynamic analysis tools~\cite{2018contractfuzzer, atzei2017survey, nguyen2020sfuzz,
echidna, wustholz2020harvey, wang2020oracle, liu2020modcon} were proposed to detect smart contract
vulnerabilities through fuzzing and model-based testing.

There is a substantial body of work focusing on the detection and exploitation of DeFi vulnerabilities. \citeauthor{qin2021attacking}~\cite{qin2021attacking} and \citeauthor{zhou2021just}~\cite{zhou2021just} manually formulated vulnerabilities such as \textit{oracle price manipulation} and \textit{arbitrage} into an optimization problem. \camera{FlashSyn~\cite{chen2022flashsyn} proposed a framework that automatically formulates flash loan attacks as a synthesis problem, by approximating the functions of DeFi protocols.}
\citeauthor{wu2021defiranger}~\cite{wu2021defiranger} identified several \textit{oracle price
manipulation} patterns from on-chain transaction data to detect real-world attacks while
\citeauthor{kong2023defitainter}~\cite{kong2023defitainter} detects price manipulation
vulnerabilities in DeFi applications through inter-contract taint analysis.
Also, \citeauthor{gudgeon2020decentralized}~\cite{gudgeon2020decentralized} showcased how to
use \textit{flashloan} to conduct \textit{governance attack}.
\citeauthor{baum2022sok}~\cite{baum2022sok} surveyed the state-of-the-art mitigation techniques for
\textit{front-running} in DeFi.
Interestingly, several works~\cite{xue2022preventing, deng2023robust, zhang2023your} explored
front-running as a defense mechanism against smart contract exploits.
Different from the over-generalized security patterns used by the existing tools,
our invariant guards capture the subtle semantic constraints of specific smart contracts.


\noindent
\textbf{Runtime Verification and Validation.} Runtime verification has been used for validation purpose where the runtime checks are on properties 
from users' expectation rather than from formal program specifications~\cite{magazzeni2017validation}.
Sereum~\cite{rodler2018sereum} is a general runtime validation framework to protect 
deployed contracts against reentrancy attacks.
Solythesis~\cite{li2020securing} provides a source-to-source compiler that compiles smart contracts with user-specified invariants to reject unexpected transactions~\cite{li2020securing}.
In contrast, {\name} is not limited to predefined attack types and has demonstrated effectiveness in mitigating a wide range of sophisticated attacks.

\section{Conclusion}\label{sec:conclusion}
In this paper, we present the first comprehensive study of the effectiveness of practical invariant guards
on preventing DeFi smart contract attacks.
Our large-scale experiments on real-world DeFi hacks demonstrate that the inferred invariant
guards are very effective in stopping the existing hacks, but some invariant guards can
be bypassed by experienced attackers. 
We also show combining multiple invariants can be more effective than individual invariants with a lower false positive rate.

\section{Data Availability} 
All benchmarks, source code, and study statistics are available at our repositories~\cite{benchmarks, artifact, study, artifactArchived}.
Additional materials of this paper are available in 
the website~\cite{website}.

\section*{Acknowledgement}
This work was supported by MITACS Accelerate program, NSERC Postdoctoral Fellowship, Singapore
Ministry of Education Academic Research Fund Tier 1 (RG12/23), and Nanyang Technological University
Centre in Computational Technologies for Finance (NTU-CCTF).

\newpage
\bibliographystyle{ACM-Reference-Format}
\bibliography{main}



\end{document}